\documentclass[a4paper,12pt]{article}
\usepackage{jheppub} 
\usepackage{braket}
\usepackage{makecell}
\usepackage{diagbox}
\usepackage{subfigure}
\usepackage{extarrows}
\usepackage{mathrsfs}
\usepackage{amsthm, amsmath}
\usepackage{amssymb}
\usepackage{mathdots}

\newcommand\td{\text{d}}
\newcommand{\p}{\partial}

\def\>{\rangle} \def\<{\langle}
\title{\boldmath BMS$_3$ modules from path integrals on the coadjoint orbits} 

\author[a]{Pujian Mao}
\author[b]{,\,Xin-Cheng Mao}

\affiliation[a]{ Center for Joint Quantum Studies, Department of Physics,\\ School of Science, Tianjin University, 135 Yaguan Road, Tianjin 300350, China}
\affiliation[b]{School of Physics, Peking University, \\No.5 Yiheyuan Rd, Beijing 100871, P.~R.~China}

\emailAdd{pjmao@tju.edu.cn,\, maoxc1120@stu.pku.edu.cn}

\abstract{
In this paper, we provide a systematic construction of quantum modules associated with coadjoint orbits of asymptotic symmetry groups through the path integral quantization at one-loop order. In particular, for the orbits with Hamiltonians unbounded from below, we show that an appropriate choice of integration contour defines a convergent Euclidean half-line path integral. The reference state associate with the orbit is determined by the path integral of the geometric action on the corresponding coadjoint orbit. The coadjoint module is therefore spanned by the reference state and its descendants. We apply this formalism to the Virasoro and BMS$_3$ groups and derive the coadjoint modules corresponding to all constant orbits. We also compute the characters of all constant Virasoro and BMS$_3$ coadjoint orbits by counting the states within the modules, and the results agree with the corresponding computations from the path integral of the geometric action with a periodic Euclidean direction.
}

\begin{document}
\maketitle
\flushbottom

\section{Introduction}

When a physical system is formulated in spacetime with boundaries, boundary effects typically arise. These may either correspond to modifications of the bulk field spectrum, such as the Casimir effect \cite{Casimir:1948dh}, or to physical degrees of freedom localized on the boundary only which are referred to as boundary degrees of freedom or edge modes \cite{Witten:1988hf,Wen:1989iv}. In gravitational theories, such boundary degrees of freedom have been intensively investigated from the perspective of asymptotic symmetry group (ASG), namely the group of symmetries acting nontrivially at the boundary. 

In three dimensions, Einstein gravity has no propagating degrees of freedom (DOF). Hence, the physical DOF is entirely encoded in boundary (global) modes associated with asymptotic symmetries. Accordingly, the gravitational solution space (classical phase space) \cite{Banados:1998gg,Barnich:2010eb} for 3D pure gravity, with or without a cosmological constant, can be classified by the corresponding ASG coadjoint orbit \cite{Nakatsu:1999wt, Navarro-Salas:1999ejl, Witten:1987ty, Garbarz:2014kaa, Barnich:2014zoa, Barnich:2015uva, Sheikh-Jabbari:2016unm}. These orbits are symplectic manifolds equipped with the canonical Kirillov--Kostant (KK) symplectic form \cite{kirillov2012elements, kirillov2025lectures}. Moreover, the three-dimensional gravitational action can be rewritten, via the Chern--Simons formulation, as a geometric action on the corresponding coadjoint orbits \cite{Cotler:2018zff, Merbis:2019wgk}. Geometric quantization of coadjoint orbits provides a representation (or module) of the corresponding symmetry algebra \cite{kirillov2012elements}. In the gravitational context, the states in the ASG coadjoint module are expected to describe the gravitational saddle and its perturbations: the saddle itself is identified with the reference state, while the perturbations are generated by the action of the symmetry algebra and can thus be viewed as its descendants.

The three-dimensional Anti-de-Sitter (AdS$_3$) cases provide a particularly clear substantiation of this picture. The ASG surface charges of AdS$_3$ under the Brown--Henneaux boundary condition generate two copies of Virasoro algebra \cite{Brown:1986nw}, which admit a holographic interpretation in terms of a two-dimensional conformal field theory, referred to as AdS$_3$/CFT$_2$ correspondence \cite{Maldacena:1997re,Gubser:1998bc,Witten:1998qj}, see also reviews in \cite{Kraus:2006wn}. The Virasoro modules associated with all constant coadjoint orbits were constructed in \cite{Witten:1987ty}. For each such orbit, the resulting module contains a highest-weight state and therefore gives rise to a highest-weight representation (HWR) of the Virasoro algebra. Thus, the Virasoro coadjoint modules for constant orbits\footnote{Here, the term ``constant orbits'' refers to orbits that admit a constant representative. For such orbits, the constant representative are chosen as the reference point. For convenience, throughout this paper we refer to these as constant orbits, while calling orbits that contain no constant representative non-constant orbits.} are precisely highest-weight modules. In particular, the pure AdS$_3$ saddle corresponds to the highest-weight vacuum state \cite{Strominger:1997eq}, and the character of the corresponding Virasoro HWR reproduces the one-loop partition function of pure AdS$_3$ gravity \cite{Witten:2007kt,Maloney:2007ud,Yin:2007gv,Giombi:2008vd,David:2009xg}. Likewise, a BTZ black hole corresponds to a primary state \cite{Strominger:1997eq}, and related descriptions of BTZ microstates in terms of boundary symmetry excitations have been explored in \cite{Afshar:2017okz}.

In the case of three-dimensional flat gravity, the ASG at null infinities is the BMS$_3$ group \cite{Ashtekar:1993ds, Ashtekar:1996cd, Barnich:2006av, Compere:2017knf}, given by the semi-direct product of supertranslations and superrotations. Upon geometric quantization, BMS$_3$ coadjoint orbits give rise to induced representations built from representations of the little group that leaves the supermomentum invariant \cite{Barnich:2015uva}. The one-loop gravitational partition function around Minkowski spacetime \cite{Barnich:2015mui} matches the character of the induced vacuum representation of BMS$_3$ algebra \cite{Oblak:2015sea}. This remarkable correspondence motivates the investigation of BMS$_3$ modules associated with various coadjoint orbits.

Naturally, one can expect to build BMS$_3$ modules following the construction of Virasoro coadjoint modules \cite{Witten:1987ty}. In the Virasoro case, the representation is specified by a single highest weight, and the construction requires only quotienting out the redundant directions associated with the stabilizer of the orbit. For BMS$_3$, however, the situation is more subtle: the coadjoint orbit is characterized by both the supermomentum and the super-angular momentum, whose dynamics are intertwined by the semidirect-product structure of the BMS$_3$ group.

Alternatively, the relation between coadjoint orbits and induced representations \cite{Barnich:2015uva} suggests to apply the standard Wigner construction of induced representations from little-group representations \cite{Wigner:1939cj}. Previous works have constructed specific BMS$_3$ induced modules, in particular the massive and vacuum representations \cite{Barnich:2014kra, Campoleoni:2016vsh}. Massive representations, with positive mass, are induced from the spin-$s$ representations $R(\theta) = e^{is \theta}$ of the little group $U(1)$, while vacuum representations, with vanishing mass and angular momentum, are induced from the trivial representation of $SL(2, \mathbb R)$. Subsequent work \cite{Garbarz:2015lua} computed the characters for some of the BMS$_3$ constant coadjoint orbits with Hamiltonians bounded from below, and found that they exactly coincide with the characters of the massive and vacuum induced representations \cite{Oblak:2015sea}. This confirms that the massive and vacuum induced modules correspond to BMS$_3$ coadjoint modules. 

Nevertheless, the remaining BMS$_3$ constant orbits are also of interest, including those for which the Hamiltonian is unbounded from below, since they belong to distinct sectors of the gravitational phase space \cite{Garbarz:2014kaa, Barnich:2015uva}. However, the corresponding construction of modules for such orbits using the standard Wigner procedure is not straightforward. Notably, as revealed in \cite{AndradeeSilva:2023cci, AndradeeSilva:2023okr}, not every induced representation corresponds to a coadjoint orbit. Thus, although the Wigner construction in principle provides a way to enumerate BMS$_3$ induced representations, it does not by itself determine which of them correspond to coadjoint orbits. Moreover, characters cannot uniquely distinguish representations of infinite-dimensional groups. A recent example \cite{Chen:2025fcc} shows that both HWR and induced representations can reproduce the same gravitational partition function, despite they are quite different representations. These observations motivate a more direct approach based on a systematic procedure that starts from a given BMS$_3$ coadjoint orbit and directly determines the corresponding module.

In this work, we construct coadjoint modules for all constant orbits of Virasoro and BMS$_3$ groups based on the Euclidean half-line path integral of the geometric action on the corresponding coadjoint orbit. At one-loop order, the path integral can always be written as a series of convergent Gaussian integral by an appropriate choice of integration contour regardless of whether the Hamiltonian is bounded from below, since such a contour always exists for any quadratic form. Given such a convergent definition, the path integral quantization determines the annihilation conditions of the reference state. Moreover, since the stabilizer acts trivially on the reference point of the coadjoint orbit, its DOFs should therefore be quotiented out in the path integral as gauge redundancies. Consequently, the generators in the stabilizer should not appear in the module. The remaining ASG generators generate independent excitations, and thereby determine the descendants of the reference state. 

We use this approach to obtain the Virasoro and BMS$_3$ coadjoint modules for all constant orbits. For the Virasoro case, our constructions of the coadjoint modules reproduce the results of \cite{Witten:1987ty}. For the BMS$_3$ case, our results covers previously known massive and vacuum modules \cite{Barnich:2014kra,Oblak:2015sea}. In the sectors where the Hamiltonian is bounded from below, we obtain a new module with vanishing mass but non-vanishing angular momentum, which is induced from a non-trivial representation of the little group $SL(2, \mathbb R)$. Importantly, we also construct the BMS$_3$ coadjoint modules from the constant orbits with Hamiltonian unbounded from below.

Once the module is known, the character can be obtained by directly counting the number of states within the module. Alternatively, the path integral of the geometric action with a periodic Euclidean direction naturally implements a trace structure and therefore yields a path-integral formulation of the character. Such path-integral character formula has a long history of investigations and applications, see, e.g., in \cite{Perret:1990bc, Alekseev:2018pbv, Alekseev:2020jja, Alekseev:1988ce, Alekseev:1988vx, Alekseev:1990mp, Aratyn:1990dj, Barnich:2017jgw,Cotler:2018zff,Merbis:2019wgk, Cotler:2024cia}, together with related studies of their one-loop exactness \cite{Stanford:2017thb,Cotler:2018zff,Merbis:2019wgk,Cotler:2024cia,Simon:2024dwm,Maxfield:2026ynw}. As a consistency check, we compute the characters of all constant Virasoro and BMS$_3$ coadjoint orbits using both the path-integral and state-counting approaches. The two methods give consistent results for all cases considered.

The paper is organized as follows. In section \ref{sec:orbit}, we give a brief review of coadjoint orbit of a Lie group. In section \ref{sec:orbitPI}, we provide the generic framework of path integral quantization on coadjoint orbits. In section \ref{sec:Virasoro}, we construct the Virasoro coadjoint modules for all constant orbits through the half-line path integral. We then perform the parallel analysis for the BMS$_3$ case and discuss their relationship with flat gravity in section \ref{sec:bms}. In section \ref{sec:conclusion}, we summarize our results and discuss future directions. In appendix \ref{sec:central}, we detail the coadjoint orbits for centrally extended case.

\section{Generic discussion on coadjoint orbit} \label{sec:orbit}

In this section, we briefly review the coadjoint representation, following \cite{kirillov2012elements,kirillov2025lectures}; see also \cite{Oblak:2016eij} for related reviews. We begin with the adjoint representation of a Lie group $G$ with Lie algebra $\mathscr G$, and then introduce the coadjoint representation, symplectic form, geometric action, and character formula. A central extension amounts to considering the coadjoint representation of the enlarged group $\widehat{G}=G\ltimes\mathbb{R}$ with Lie algebra $\widehat{\mathscr G}=\mathscr G\oplus\mathbb{R}$. Thus, the discussion below applies directly to centrally extended groups as well, and we review the relevant details in appendix \ref{sec:central}.

The adjoint representation is given by
\begin{equation}
    \text{Ad}_g v = g v g^{-1}, \qquad \text{ad}_u v = \p_t \left[ \text{Ad}_{g_t} v \right] \big|_{t=0} = [u,v],
\end{equation}
where $[u,v]$ is the Lie bracket and $g_t \in G$ is the integral curve of the vector $u=\p_t g_t |_{t=0} \in \mathscr G$. Any Lie algebra element can be expanded through the basis $t_i$ by $u = u^i t_i$, where $[t_i, t_j] = f^k_{ij} t_k$ and $f^k_{ij}$ is the structure constant. The space of linear functionals on $\mathscr G$ forms its dual space $\mathscr G^*$, which defines the inner product structure
\begin{equation}
    b: \mathscr G \mapsto \mathscr G^* : v \mapsto b(v)=\<b,v\>.
\end{equation}
The coadjoint vector $b$ can be expanded through the coordinates $b_i$ and the tangent vector $(t^i)^*$ on the orbit as $b = b_i (t^i)^*$. Note that the vector $(t^i)^*$ is dual to the adjoint element $t_i$, and their inner product is $\<(t^i)^*, t_j \> = \delta^i_j$. The coadjoint representations acting on the linear space $\mathscr G^*$ are defined as
\begin{equation} \label{eq:Ad*def}
    \<\text{Ad}^*_g b, v\> = \<b, \text{Ad}_{g^{-1}} v\>, \quad \<\text{ad}^*_u b, v\> = - \<b, \text{ad}_u v\>
\end{equation}

The coadjoint orbit of $b_0 \in \mathscr G^*$ is defined as
\begin{equation}
    W_{b_0} = \left\{ \text{Ad}^*_g b_0 \big| g \in G \right\} \simeq G/\text{Stab}(b_0),
\end{equation}
with the stabilizer of the element $b_0$ given by
\begin{equation}
    \text{Stab} (b_0) = \left\{ g \big| \text{Ad}^*_g b_0 = b_0, ~ g \in G \right\}, \quad \mathfrak{stab} (b_0) = \left\{ u \big| \text{ad}^*_u b_0 = 0, ~ u \in \mathscr G \right\}.
\end{equation}
Since every point on the orbit can be linked to $b_0$ through finite transformation $b = \text{Ad}^*_g b_0$, its variation in terms of $g$ is then
\begin{equation} \label{eq:delub}
    \delta \< \text{Ad}^*_g b_0, v\> = \< \text{ad}^*_u (\text{Ad}^*_g b_0), v \>, \qquad u = \delta g g^{-1}, 
\end{equation}
where $u = \delta g g^{-1}$ is known as the right-invariant Maurer--Cartan (MC) form.  

The KK simplectic form $\Omega_b$ of the tangent space of the coadjoint orbit is
\begin{equation} \label{eq:Omega}
    \Omega_b = \<b, \delta_u u\> = \delta_u \alpha_b, \qquad \alpha_b = - \<b, u \> = - \< \text{Ad}^*_g b_0, \dot g g^{-1} \> \td t,
\end{equation}
where $u$ is the MC-form of $b$ given in \eqref{eq:delub}, and $\alpha_b$ is the symplectic potential. Inserting two vectors $\text{ad}^*_v b, \text{ad}^*_{v'} b$ tangent to the orbit $W_b$, one obtains
\begin{equation} \label{eq:Omegab}
    \Omega_b (\text{ad}^*_v b, \text{ad}^*_{v'} b ) = \<b, [v,v'] \> = \Omega_{b, ij} v^i v'^j, \quad \Omega_{b, ij} = f_{ij}^k b_k.
\end{equation}
Therefore, the Poisson bracket between adjoint vectors is defined through the inverse of the symplectic matrix $(\Omega_b^{-1})_{ij}$. In contrast, the Poisson bracket between coadjoint vectors is defined through the symplectic matrix $\Omega_{b, ij}$ itself, instead of its inverse. Moreover, the conserved charges and their Poisson bracket are then given by \cite{Witten:1987ty}
\begin{equation} \label{eq:charge}
    \Phi_v (b) = \<b, v\> = b_i v^i, \quad \{\Phi_v, \Phi_{v'}\} = \Phi_{[v,v']} = \Omega_b (\text{ad}^*_v b, \text{ad}^*_{v'} b).
\end{equation}
Therefore, the charges are expanded by the basis $b_i = \<b, t_i\>$ and one can show that the Poisson bracket $\{b_i, b_j\} = \Omega_{b, ij} = f_{ij}^k b_k$ repreduces the Lie algebra of $[t_i, t_j]$.

Through the Darboux coordinates $(q_s, p_s)$ in the phase space, the KK-form is diagonalized as blocks
\begin{equation} \label{eq:Darboux}
    \Omega_b = \delta q_s \land \delta p_s = \delta \alpha_b, \qquad \alpha_b = - \< \text{Ad}^*_g b_0, \delta g g^{-1} \> = -p_s \delta q_s.
\end{equation}
The Poisson bracket for Darboux coordinates is $\{q_s, p_{s'}\} = \delta_{s, s'}$. The symplectic potential $\alpha_b$ provides the kinetic term of the action of the phase space, which is normally referred to as the geometric action,
\begin{equation} \label{eq:kinetic}
    A = \int_\Sigma \Omega_b = \int_{\gamma(t)} \alpha_b = - \int \< \text{Ad}^*_g b_0, \dot g g^{-1} \> \td t = - \int \td t ~ p_s \dot q_s,
\end{equation}
where $\Sigma$ is a two-surface in the phase space with the path $\gamma$ being its boundary.

\section{Path integral quantization on coadjoint orbit} \label{sec:orbitPI}

In this section, we present the general formulation of constructing the module associated with a given coadjoint orbit via the path integral quantization. Our analysis is performed perturbatively and restricted to one-loop order. As discussed in section \ref{sec:orbit}, a coadjoint orbit is a classical phase space. Its quantization therefore gives rise to a Hilbert space of quantum mechanics. In particular, at the quadratic order in perturbation theory, the path-integral quantization leads to a quantum-mechanical system. In the coordinate representation associated with Darboux coordinates, the transition amplitude can be expressed in terms of a path integral as
\begin{equation}
    \begin{split}
        \bra{q^{(N)}} e^{-i \hat{\bar H} \beta} \ket{q^{(0)}} & = \int \prod^{N-1}_{j=1} Dq^{(j)} \prod^{N-1}_{n=0} \bra{q^{(n+1)}} e^{-i \hat{\bar H} \frac{\beta}{N}} \ket{q^{(n)}} \\
        & = \int_{q(t=-\beta)=q^{(0)}}^{q(t=0)=q^{(N)}} Dq \int Dp e^{iS^{(\beta)}}
    \end{split}
\end{equation}
where $S^{(\beta)}$ is the action deformed from the geometric action $A$ defined in \eqref{eq:kinetic} by the Hamiltonian $\bar H$ with chemical potential $a_v/ \beta$
\begin{equation}
    S^{(\beta)} = \int^0_{-\beta} (p_s \dot q_s - \bar H) \td t = -A - \int^0_{-\beta} \bar H \td t, \qquad \bar H = H + \sum_v \frac{a_v}{\beta} \Phi_v (b),
\end{equation}
and $H, \Phi_v$ are the generators of time-direction and $a_v$-direction given by the inner products \eqref{eq:charge}, and $\hat{\bar H}$ in the transition amplitude is the operator corresponding to $\bar H$. Moreover, we used the abbreviations $q = (q_1, \cdots, q_s), p = (p_1, \cdots, p_s)$ for the Darboux coordinates, used $q^{(n)}, p^{(n)}$ to denote the value of $q, p$ at $t=(n-N)\beta/N$, inserted the identities $\int Dq^{(n)} \ket{q^{(n)}} \bra{q^{(n)}} = \int Dp^{(n)} \ket{p^{(n)}} \bra{p^{(n)}} = 1$, and used the inner product $\bra{q^{(n)}} p^{(m)}\> = e^{ip_s^{(m)} q_s^{(n)}}$ and $q^{(n+1)}_s-q^{(n)}_s = \dot q^{(n)}_s \beta/N$. Here, the integral measure $DpDq= \prod_s Dp_s Dq_s$ counts the degrees of freedom in the phase space of the coadjoint orbit, which can be formally rewritten as $\frac{Db}{\text{Stab}(b_0)}$, where the stabilizer of the orbit is modded out since it is the gauge redundancy. 

With analytical continuation $\beta \rightarrow -i\beta$, the above formula can be rewritten under the Euclidean signature as
\begin{equation} \label{eq:PIamplitude}
    \begin{split}
        & \bra{q^{(N)}} e^{- \hat{\tilde H} \beta} \ket{q^{(0)}} =\int_{q(t=-\beta)=q^{(0)}}^{q(t=0)=q^{(N)}} Dq \int Dp e^{- \tilde S_E^{(\beta)}}, \\
        & \tilde S_E^{(\beta)} = iA + \int^0_{-\beta} \tilde H \td t, \quad \tilde H = H + i \sum_v \frac{a_v}{\beta} \Phi_v (b).
    \end{split}
\end{equation}
Note that the state $\ket{q}$ can be expanded by the eigenbasis $\ket{n}$ of $\hat{\tilde H}$, with $\hat{\tilde H} \ket{n}=\tilde E_n\ket{n}$. When Re$\hat{\tilde H}$ is bounded from below, the contribution with the smallest real part of the eigenvalue, Re$\tilde E_0$, dominates in the large-$\beta$ limit. Thus, by taking $\beta\to\infty$, one can project onto the ground state
\begin{equation}
    e^{- \hat{\tilde H} \beta} \ket{q} = \sum_n e^{- \tilde E_n \beta} \ket{n} \<n \ket{q} \underset{\beta \to \infty}{\longrightarrow} e^{- \tilde E_0 \beta} \<0 \ket{q} \ket{0}.
\end{equation}
Therefore, one can define the ground state on this quantized orbit by half-line path integral
\begin{equation} \label{eq:halflinePI}
    \bra{q^{(N)}} b_0\> = \int^{q^{(N)}} Dq \int Dp e^{- \tilde S_E^{(\infty)}},
\end{equation}
where the ground state $\ket{0}$ is actually regarded as the reference state $\ket{b_0} = e^{- \tilde E_0 \infty} \<0 \ket{q} \ket{0}$ within the module corresponding to this orbit. The excitation states in the module are descendants of $\ket{b_0}$. Note that in this case the Darboux coordinates are chosen as real numbers.

However, there are in general many coadjoint orbits for which the Hamiltonian is unbounded from below. In such cases, the $\beta \to \infty$ limit in \eqref{eq:PIamplitude} cannot extract the ground state due to the existence of negative modes in the Hamiltonian, corresponding to a wrong-sign Gaussian integral at the one-loop level of the path integral in \eqref{eq:PIamplitude}. Consequently, defining the path integral in the same way as for orbits with a Hamiltonian bounded from below does not yield \eqref{eq:halflinePI} for the orbits with unbounded Hamiltonians. Fortunately, for any quadratic form, one can always choose an appropriate integration contour such that the resulting Gaussian integral has the correct sign. This provides a convergent definition of path integral at one-loop order regardless of whether the Hamiltonian is bounded from below. 

The path integral quantization then gives the annihilation conditions of the reference states, which arise from two different mechanisms in the half-line path integral. On the one hand, the contribution of Re$\tilde H = H$ in the half-line path integral under an appropriate contour determines that \eqref{eq:PIamplitude} and \eqref{eq:halflinePI} are convergent for arbitrary $q^{(N)}$. Following the same analysis as in \cite{Chen:2025acl, Chen:2025eeh}, this implies the following regularity conditions on the reference state
\begin{equation} \label{eq:regular}
    q \ket{b_0}, p\ket{b_0} = \text{finite, ~~ when}: t \to -\infty.
\end{equation}
On the other hand, the contribution of $\Phi_v$ in the path integral might also generate other annihilation conditions, which cannot be uniquely determined by the half-line regularity condition since $\Phi_v$ enter the path integral as phase factors rather than exponential suppression. Therefore, one has to specify which modes are annihilated by polarization. Moreover, since the stabilizer DOFs have already been quotiented out in the definition of the path integral, the corresponding generators should likewise be absent from the resulting module, either as null states or by annihilating the reference state directly. These annihilation conditions determine which generators annihilate the reference state, and hence should not appear in the resulting module.
 
After deriving the module, it is easy to calculate the character for the corresponding coadjoint orbit by counting the states within the module. Equivalently, the character can also be derived through the well-defined Euclidean path integral \eqref{eq:PIamplitude} by introducing the periodic condition
\begin{equation} \label{eq:chiHEucl}
    \text{Tr} e^{-\tilde H \beta} = \int Dq^{(N)} Dq^{(0)} \delta (q^{(0)} - q^{(N)}) \bra{q^{(N)}} e^{- \hat{\tilde H} \beta} \ket{q^{(0)}} = \int_{t \sim t+\beta} DqDp e^{-\tilde S_E^{(\beta)}}.
\end{equation}

\section{Coadjoint orbits of Virasoro group} \label{sec:Virasoro}

As a consistency check of the generic framework in previous section, we first study the Virasoro case and verify that the Virasoro coadjoint module constructed through the half-line path integral on the constant orbits precisely reproduce the previously studied Virasoro HWRs \cite{Witten:1987ty}. The Virasoro algebra is generated by the central extension of the Witt algebra, generating the diffeomorphism on $S^1$. The finite transformation is denoted by $g \in \text{Diff} (S^1): \theta \mapsto g (\theta)$ satisfying $g (\theta + 2\pi) = g (\theta) + 2 \pi$. Its integral curve is parameterized by $t$. For small $t$, it can be expanded as
\begin{equation} \label{eq:gtsmall}
    g_t (\theta) = \theta + t u (\theta) + O(t^2), \qquad u (\theta + 2\pi) = u(\theta).
\end{equation}
The centrally extended Lie algebra \eqref{eq:Liec} in this case is then given by
\begin{equation}
	 \widehat{\text{ad}}_{(u, \lambda)} (v, \mu) = [(u,\lambda), (v, \mu)] = - \left(u v' - v u', \frac{1}{24 \pi} \int^{2 \pi}_0 \td \theta u''' v \right).
\end{equation}
The inner product \eqref{eq:innerc} for Virasoro case is
\begin{equation} \label{eq:innerVirac}
    \<(b, c), (u, n)\> = \<b, u\>_0 + cn, \qquad \<b, u\>_0 = \int^{2 \pi}_0 \td \theta b (\theta) u (\theta).
\end{equation}
Then, the coadjoint action \eqref{eq:Ad*gbc} for Virasoro group is derived as
\begin{equation} \label{eq:ad*b0Vir}
    \begin{split}
        & \widehat{\text{Ad}}^*_f (b (\theta) ,c) = ((\text{Ad}^*_f b) (\theta) - c S_\theta (f^{-1}), c) = ( b (F(\theta)) F'^2 (\theta) - c S_\theta (F), c), \\
        & \widehat{\text{ad}}^*_{(u, \lambda)} (b (\theta) ,c) = (\delta b, \delta c), \quad \delta b = - u b' - 2 u' b + \frac{c}{24 \pi} u''' , \quad \delta c = 0,
    \end{split}
\end{equation}
where the coadjoint representation for $c=0$ is 
\begin{equation} \label{eq:Adf*b}
    (\text{Ad}^*_f b) (f(\theta)) = \frac{b (\theta)}{(f' (\theta))^2}, \quad (\text{ad}^*_u b) (\theta) = - u (\theta) b' (\theta) - 2 u' (\theta) b (\theta).
\end{equation}
Moreover, we used $F = f^{-1} \approx \theta - t u (\theta) + O (t^2)$, and the one-cocycles $S(g), s(u)$
\begin{equation} \label{eq:Schwarz}
    S_\theta [g_t(\theta)] = \frac{1}{24 \pi} \left[ \frac{g'''_t (\theta)}{g'_t (\theta)} - \frac32 \frac{g''^2_t (\theta)}{g'^2_t (\theta)} \right], \quad s_\theta (u) = \frac{1}{24 \pi} u'''(\theta),
\end{equation}
which are derived from \eqref{eq:CS}, by applying the two-cocycle
\begin{equation} \label{eq:CVir}
    C(f,g) = -\frac{1}{48 \pi} \int^{2 \pi}_0 \td \theta \log [f' (g (\theta))] \frac{g'' (\theta)}{g' (\theta)}.
\end{equation}

The Virasoro coadjoint orbit $\mathcal{W}_{(b_0, c)}$ is classified by the stabilizer of the reference point $b_0$ satisfying $\delta b_0 = 0$, where $\delta b_0$ is defined in \eqref{eq:ad*b0Vir}. Let us first consider the constant orbits with $b'_0=0$, and the stabilizer is given by \cite{Witten:1987ty}
\begin{equation} \label{eq:StabVir}
    \mathfrak{stab} (b_0) = 
    \begin{cases}
        & u=u_0,\qquad b_0 \neq - \frac{n^2 c}{48 \pi} (\forall n \in \mathbb Z_+) \\
        & u = e^{\pm i k \theta}, \text{Constant}, \qquad b_0 = - \frac{k^2 c}{48 \pi}
    \end{cases}
\end{equation}
where $u_0$ is a constant. The stabilizer can also be derived by considering the absence of excitation modes for the excited state $\ket{b}$, which can be observed perturbatively from the infinitesimal version of the mode expansions $b = \sum_{n \in \mathbb Z} b_n e^{-in \theta}, u = \sum_{n \in \mathbb Z} u_n e^{-in \theta}$ on constant orbits
\begin{equation}
    b (\theta) = b_0 + \delta b_0 = b_0 - 2i t \sum_{n \in \mathbb Z} n \beta_n u_n e^{i n \theta} + O(t^2), \quad \beta_n = b_0 + \frac{c n^2}{48 \pi},
\end{equation}
with infinitesimal excitation $t u_n$ around the reference point $b_0$. When $b_0 = - \frac{k^2 c}{48 \pi} (k \in \mathbb Z_+)$, the modes $u=u_{\pm n} e^{\pm i k \theta}, u_0$ are absent, since the coefficient $n \beta_n$ vanishes at $n =0, \pm k$. When $b_0 \neq - \frac{k^2 c}{48 \pi}, \forall k \in \mathbb Z_+$, the $u=u_0$ mode is absent, since the coefficient $n \beta_n$ only vanishes at $n =0$. The non-constant orbits were firstly discussed by considering perturbations around constant orbits in \cite{Witten:1987ty}, and were investigated non-perturbatively using Hill's equation as in \cite{Balog:1997zz, Sheikh-Jabbari:2014nya}. In this paper, we only focus on constant orbits.

The symplectic form \eqref{eq:Omegab} for Virasoro group is perturbatively obtained as
\begin{equation} \label{eq:Virasymp}
	\left\< \widehat{\text{Ad}}^*_f (b_0, c), [(u, 0), (v, 0)] \right\> = \sum_{n,m \in \mathbb Z} \Omega_{b,nm} u_n v_m, \quad \Omega_{b, nm} \approx - 4 \pi i n \beta_n \delta_{n+m}.
\end{equation}
The Virasoro generators can also be perturbatively computed as
\begin{equation} \label{eq:L+-n}
    \mathcal{L}_k = \left\< \widehat{\text{Ad}}^*_f (b_0, c), (e^{- ik \theta}, 0) \right\> \approx 
    \begin{cases}
        & 2 \pi b_0 + 4 \pi \sum_{n =1}^\infty F_n F_{-n} n^2 \beta_n, \quad k =0 \\
        & 4 \pi i k F_{-k} \beta_k, \qquad k \neq 0
    \end{cases};
\end{equation}
where the coefficients $F_n$ come from the mode expansion 
\begin{equation} \label{eq:Fexpansion}
    F(\theta, t) = f^{-1} = \theta + \sum_{n \in \mathbb Z} e^{-in \theta} F_n (t). 
\end{equation}
The Poisson bracket of $F_n$ can be derived by
\begin{equation} \label{eq:FFPoisonVir}
    \{F_n, F_m\} = (\Omega_{b, nm})^{-1} = \frac{i}{4 \pi n \beta_n} \delta_{n+m}, \qquad n \beta_n \neq 0.
\end{equation}
From the definition in \eqref{eq:L+-n} and the Poisson bracket between charges \eqref{eq:PoissonPhi}, one can check that the generators $\mathcal{L}_n$ satisfy the Virasoro algebra
\begin{equation} \label{eq:VirasorA}
    [\mathcal{L}_n, \mathcal{L}_m] = i \{\mathcal{L}_n, \mathcal{L}_m\} = (n-m) \mathcal{L}_{n+m} + \frac{c}{12} n^3 \delta_{n+m,0}.
\end{equation}
Moreover, the geometric action \eqref{eq:geoc} of the Virasoro group is
\begin{equation} \label{eq:AVir}
    A = -\left\<\widehat{\text{Ad}}^*_f (b_0, c), (u, m_u) \right\> \approx 4 \pi i \int^0_{-\beta} \td t \sum_{n =1}^\infty F_{-n} \dot F_n n \beta_n,
\end{equation}
where the prime and overhead dot denote $\p_\theta$ and $\p_t$, and the MC form \eqref{eq:umu} is given by \cite{Alekseev:1988ce, Aratyn:1990dj}
\begin{equation} \label{eq:muVir}
    \begin{split}
        u (\theta) & = (\delta f) (f^{-1} (\theta)) = \delta [f (f^{-1} (\theta))] - f'(f^{-1} (\theta)) \delta f^{-1} = - \frac{\delta f^{-1} (\theta)}{(f^{-1})' (\theta)}, \\
        m_u & = \delta_g C(g, F)|_{g=F^{-1}=f} = \frac{-1}{48 \pi} \int^{2 \pi}_0 \td \theta \frac{\delta F}{F'} \left( \frac{F'''}{F'} - \frac{F''^2}{F'^2} \right).
    \end{split}
\end{equation}

From the definitions of Virasoro generators \eqref{eq:L+-n}, one concludes that the two types of stabilizers of constant orbits derived in \eqref{eq:StabVir} are 
\begin{equation}
    u(1) = \{(1, 0)\} = \{\mathcal{L}_0\}, \quad sl^{(k)} (2, \mathbb R) = \{(e^{\pm ik \theta}, 0), (1,0)\} = \{\mathcal{L}_0, \mathcal{L}_{\pm k}\}.
\end{equation}
We will then separately discuss their coadjoint modules, and characters from both the state-counting in the modules and the path integral quantization.

\subsection{The orbits with stabilizer $u(1)$}

In this case, we have $\beta_n \neq 0 (\forall n \in \mathbb Z_+)$. We can choose the Darboux coordinates that diagonalize $\Omega_{b, nm}$ into blocks as
\begin{equation} \label{eq:DarVir}
    q_n = \sqrt{2 \pi \beta_n} (F_n + F_{-n}), \qquad p_n = n \sqrt{2 \pi \beta_n} i (F_n - F_{-n}), \qquad \forall n>0.
\end{equation}
One can easily check that $\{q_n, p_m\} = \delta_{m,n}$ using the Poisson bracket \eqref{eq:FFPoisonVir}. In terms of these Darboux coordinates, the Hamiltonian $\mathcal{L}_0$ and geometric action can be rewritten as
\begin{equation} \label{eq:AL0u(1)}
    \mathcal{L}_0 \approx 2 \pi b_0 + \sum_{n = 1}^\infty \frac{1}{2} (p_n^2 + n^2 q_n^2),\qquad A \approx -\sum_{n =1}^\infty \int^\beta_0 p_n \dot q_n \td t.
\end{equation}
Based on the discussion in section \ref{sec:orbitPI}, the transition amplitude on these types of orbits is given by
\begin{equation}
    \begin{split}
        \bra{q^{(N)}} e^{- \beta \hat{\mathcal{L}}_0} \ket{q^{(0)}} & = e^{- 2 \pi b_0 \beta} \prod_{n =1}^\infty \int Dp_n Dq_n e^{\int^0_{-\beta} \td t (i p_n \dot q_n - \frac{p_n^2 + n^2 q_n^2}{2})} \\
        & = e^{- 2 \pi b_0 \beta} \prod_{n =1}^\infty \frac{e^{-\frac{n}{2 \sinh (n\beta)} [((q^{(N)})^2+ (q^{(0)})^2) \cosh (n \beta) - 2 q^{(N)} q^{(0)}]}}{\sqrt{2 \pi \sinh (n \beta)}},
    \end{split}
\end{equation}
which is nothing but a collection of path integral of harmonic oscillators \cite{feynman1966quantum, cohen1998path}.

The path integral is defined by choosing an integration contour in the complex $F_n$ plane. For orbits with a Hamiltonian bounded from below, namely $\beta_1>0$, one usually requires the $F (\theta)$ in \eqref{eq:Fexpansion} to be real, and the integration contour satisfies 
\begin{equation}
    F_n^* = F_{-n}, \qquad \forall n \in \mathbb Z_+.
\end{equation}
On the other hand, an orbit with a Hamiltonian unbounded from below can be characterized by $\beta_{n_0}<0<\beta_{n_0+1} (n_0\in\mathbb Z_+)$. In this case, a convergent definition of the path integral is obtained by choosing a contour satisfying
\begin{equation}
    F_n^* = \begin{cases}
        F_{-n}, & n >n_0 \\
        -F_{-n}, & 1 \leq n \leq n_0
    \end{cases}.
\end{equation}
A relevant local reversal of the reality condition for negative Virasoro modes was proposed in \cite{Alekseev:2020jja} in connection with a Lefschetz-thimble interpretation. Here, we have defaulted to the $c>0$ case, and the analysis of $c<0$ case is similar, which will not be reiterated here, nor in the later sections.

The reference state $\ket{b_0}$ is prepared by the corresponding Euclidean half-line path integral on the orbit
\begin{equation}
    \prod_{n =1}^\infty \int Dp_n Dq_n e^{\int^0_{-\infty} \td t (i p_n \dot q_n - \frac{p_n^2 + n^2 q_n^2}{2})},
\end{equation}
with the Darboux coordinates defined in \eqref{eq:DarVir}. The Euclidean path integral of quadratic form will be localized to the configuration that satisfies the equation of motion
\begin{equation}
    p_n = i \dot q_n, \qquad \ddot q_n - n^2 q_n = 0,
\end{equation}
the solution of which is
\begin{equation}
    q_n = a_n^\dagger e^{n t} + a_n e^{-n t}, \qquad p_n = i n (a_n^\dagger e^{n t} - a_n e^{-n t}),
\end{equation}
or reformulated as follows using \eqref{eq:DarVir},
\begin{equation}
    \begin{split}
        a_n e^{-nt} & = \frac12 \left(q_n - \frac{p_n}{in} \right) = \sqrt{2\pi \beta_n} F_{-n},\quad a_n^\dagger e^{nt} = \frac12 \left(q_n + \frac{p_n}{in} \right) = \sqrt{2\pi \beta_n} F_n.
    \end{split}
\end{equation}
Note that the $e^{-n t}$ mode is divergent at $t \to -\infty$ boundary. Consequently, the regularity condition \eqref{eq:regular} originates from the path integral quantization directly gives the annihilation condition of harmonic oscillators
\begin{equation} \label{eq:an|0>=0}
    F_{-n} \ket{b_0} = a_n \ket{b_0} = 0 \qquad (\forall n \in \mathbb Z_+).
\end{equation}

By definition, the path integrals on the orbits remove the degree of freedom from the stabilizer. Therefore, the action of $\mathcal{L}_0$ on the reference state does not provide an independent excitation in the resulting module. This can be seen more precisely from \eqref{eq:L+-n}. At the quantum level, all products of $F_n$ must be normal ordered, denoted by $:\cdots:$, with modes $F_n$ carrying smaller indices placed to the right. Using the annihilation conditions in \eqref{eq:an|0>=0}, we have $:F_n F_{-n}: \ket{b_0} = 0$ for $n \neq 0$. The action of $\mathcal{L}_0$ on the reference state then contains only a constant term, i.e., $\mathcal{L}_0 \ket{b_0} = 2 \pi b_0 \ket{b_0}$. Moreover, using the relation $\mathcal{L}_n \propto F_{-n} (n \neq 0)$ as shown in \eqref{eq:L+-n}, the annihilation conditions \eqref{eq:an|0>=0} can also be reformulated by $\mathcal{L}_n$. Consequently, the full definition of the reference state is
\begin{equation} \label{eq:Ln|0>Vir}
    \mathcal{L}_0 \ket{b_0} = 2 \pi b_0 \ket{b_0}, \qquad \mathcal{L}_n \ket{b_0} = 0, \qquad (n \in \mathbb Z_+).
\end{equation}
The module is therefore spanned by
\begin{equation} \label{eq:Ldescendant}
    :\mathcal{L}_{-n_1} \cdots \mathcal{L}_{-n_i}: \ket{b_0},~~ (i = 0, 1, 2, \cdots); ~ n_1, \cdots, n_i \in \mathbb Z_+.
\end{equation}
In particular, $i=0$ stands for the reference state $\ket{b_0}$. Here, the normal ordering places the Virasoro generator with smaller indices to the left, since $\mathcal{L}_n \propto F_{-n}$. 

From the Virasoro algebra \eqref{eq:VirasorA}, one can easily derive that the descendants \eqref{eq:Ldescendant} are also the eigenstates of $\mathcal{L}_0$
\begin{equation}
    \mathcal{L}_0 :\mathcal{L}_{-n_1} \cdots \mathcal{L}_{-n_i}: \ket{b_0} = (2 \pi b_0 + n_1 + \cdots + n_i) :\mathcal{L}_{-n_1} \cdots \mathcal{L}_{-n_i}: \ket{b_0}.
\end{equation}
Therefore, the orbits with the stabilizer $u(1)$ are HWRs, since the reference state $\ket{b_0}$ possesses the lowest eigenvalue $2 \pi b_0$ of $\mathcal{L}_0$ among all states in the module \eqref{eq:Ldescendant}. Furthermore, it is easy to calculate the character by counting the number of states in the module \eqref{eq:Ldescendant}
\begin{equation} \label{eq:S1character}
    \mathrm{Tr} e^{-L_0 \beta} = \sum_{N=0}^\infty e^{-\beta (2 \pi b_0 +N)} p(N) = \frac{q^{2 \pi b_0}}{\prod^\infty_{n=1} (1-q^n)}, \quad q = e^{-\beta},
\end{equation}
where $p(N)$ counts the distinct partitions of the integer $N$, and we used the Euler function
\begin{equation} \label{eq:Euler}
    \frac{1}{\varphi (x)} = \prod^\infty_{n = 1} \frac{1}{1-x^n} = \sum_{n = 0}^\infty p (n) x^n.
\end{equation}
Alternatively, the character \eqref{eq:S1character} can also be derived through the path integral \eqref{eq:chiHEucl}
\begin{equation}
    \mathrm{Tr} e^{-\beta \mathcal{L}_0} = q^{2 \pi b_0} \prod_{n =1}^\infty \chi^{H (m=1)}_{(n, \beta)}, 
\end{equation}
where we formally used the path integral for 1d harmonic oscillator \cite{feynman1966quantum, cohen1998path},
\begin{equation} \label{eq:1dHarOPI}
    \chi^{H (m)}_{(\omega, \beta)} = e^{\omega \beta/2} \int Dp Dq e^{\int^\beta_0 \td t (i p \dot q - \frac{p^2 + m^2 \omega^2 q^2}{2m})} = \frac{1}{1- e^{-\omega \beta}}.
\end{equation}
Here, the vacuum zero-point energies $e^{-\omega \beta/2}$ contribution from the oscillators are absorbed into $b_0$, which is a renormalization of the orbit parameter $b_0$. Notice that this prescription is consistent with the previous state-counting computation \eqref{eq:S1character}, where the zero-point energies were removed by adopting the normal-ordering procedure.

\subsection{The orbits with stabilizer $sl^{(k)} (2, \mathbb R)$}

The reference point of the orbits with stabilizer $sl^{(k>0)} (2, \mathbb R)$ is $b_0 = - \frac{ck^2}{48 \pi}$ (namely $\beta_k = 0$), indicating that the $k$-th Darboux coordinates \eqref{eq:DarVir} vanishes, i.e., $q_k = p_k = 0$. Then, the geometric action and the Hamiltonian can be rewritten as
\begin{equation} \label{eq:AL0slk}
    A \approx -\sum_{n \in \mathbb Z_+/ \{k\}} \int^\beta_0 p_n \dot q_n \td t, \qquad \mathcal{L}_0 \approx 2 \pi b_0 + \sum_{n \in \mathbb Z_+/ \{k\}} \frac{1}{2} (p_n^2 + n^2 q_n^2).
\end{equation}
Thus, the system can again be viewed as a collection of one-dimensional harmonic oscillators with positive integer frequencies, and the analysis proceeds analogously. The only difference from the $u(1)$ case is that the oscillator with frequency $k$ is absent. Then, the transition amplitude is given by the path integral as 
\begin{equation}
    \bra{q^{(N)}} e^{- \beta \hat{\mathcal{L}}_0} \ket{q^{(0)}} = e^{- 2 \pi b_0 \beta} \prod_{n \in \mathbb Z_+/ \{k\}} \int Dp_n Dq_n e^{\int^0_{-\beta} \td t (i p_n \dot q_n - \frac{p_n^2 + n^2 q_n^2}{2})}.
\end{equation}
For the case where $k>1$, the Hamiltonian is unbounded from below, and the contour of the convergent path integral is chosen as
\begin{equation}
    F_n^* = \begin{cases}
        F_{-n}, & n >k \\
        -F_{-n}, & 1 \leq n <k
    \end{cases}.
\end{equation}
When $k=1$, the Hamiltonian has a lower bound, and the integration contour is defined as
\begin{equation}
    F_n^* = F_{-n}, \qquad n =2,3,4, \cdots.
\end{equation}

One can similarly construct the module of the orbit. Since the $k$-th harmonic oscillator is missing in this case, the regularity condition \eqref{eq:regular} of the path integral only gives
\begin{equation} \label{eq:annihislkPI}
    F_{-n} \ket{b_0} = 0, \qquad (\forall n \in \mathbb Z_+/\{k\}).
\end{equation}
Note that $\mathcal{L}_0, \mathcal{L}_{\pm k}$ should not appear in the final module, since they belong to the stabilizer. More precisely, this follows from \eqref{eq:L+-n} upon imposing the normal-ordering prescription. As before, $\ket{b_0}$ is an eigenstate of $\mathcal{L}_0$. Moreover, we need to expand $\mathcal{L}_{\pm k}$ to the next order when $\beta_k =0$,
\begin{equation}
    \mathcal{L}_{\pm k} = \frac{-c}{48 \pi} \sum_{n \in \mathbb Z} n (n \pm k) (n^2 \mp kn - k^2) :F_n F_{-n \mp k}: + O(F_n^3).
\end{equation}
We then find that every term in $\mathcal{L}_k$ contains at least one $F_{-n}$ with $n \in \mathbb Z_+/\{k\}$, which annihilates the reference state according to \eqref{eq:annihislkPI}, and is placed on the right by normal ordering. Hence, we have $\mathcal{L}_k \ket{b_0} = 0$. By contrast, $\mathcal{L}_{-k}$ contains terms involving two $F_n$ modes with positive indices, and therefore does not annihilate the reference state. Altogether, we obtain the definition of the reference state
\begin{equation}
    \begin{split}
        &\mathcal{L}_n \ket{b_0} =0 \quad (\forall n \in \mathbb Z_+), \qquad \mathcal{L}_0 \ket{b_0} = 2 \pi b_0 \ket{b_0}, \\
        & \bigg[\mathcal{L}_{-k} + \frac{3}{c \pi} \sum_{n =1}^{k-1} \frac{n^2 + kn - k^2}{n (2k-n) (n^2-k^2)} :\mathcal{L}_{-n} \mathcal{L}_{n - k}: + O(F_n^3) \bigg] \ket{b_0} = 0.
    \end{split}
\end{equation}
The latter shows that the action of $\mathcal{L}_{-k}$ does not generate an independent excitation, since it can be expressed in terms of the other Virasoro generators. The resulting Virasoro module is therefore spanned by
\begin{equation}
    :\mathcal{L}_{-n_1} \cdots \mathcal{L}_{-n_i} : \ket{b_0},~ ~ (i = 0, 1, 2, \cdots); ~ n_1, \cdots, n_i \in \mathbb Z_+ /\{k\},
\end{equation}
in agreement with the construction in \cite{Witten:1987ty}. In particular, for $k=1$, this reduces to the standard Virasoro vacuum HWR.

The character can therefore be obtained by counting the states in this module. However, there is no need to repeat the calculation; we only need to subtract the contribution of the descendants including $L_{-k}$ mode in \eqref{eq:S1character}, which is equivalent to multiplying \eqref{eq:S1character} by a factor $1-q^k$, namely
\begin{equation} \label{eq:slkcharacter}
    \mathrm{Tr} e^{-L_0 \beta} = \frac{q^{2 \pi b_0}}{\prod_{n\in \mathbb Z_+/\{k\}} (1-q^n)}.
\end{equation}
Alternatively, it can also be reproduced through the path integral by using \eqref{eq:1dHarOPI},
\begin{equation}
    \mathrm{Tr} e^{-\beta \mathcal{L}_0} = q^{2 \pi b_0} \prod_{n \in \mathbb Z_+/\{k\}} \chi^{H (m=1)}_{(n, \beta)}, 
\end{equation}
where we also absorb the zero-point energy of harmonic oscillators by renormalize the orbit parameter $b_0$.

\section{Coadjoint orbits of BMS$_3$ group}
\label{sec:bms}

The structure of BMS$_3$ group is a semidirect product of superrotations $f(\theta)$ and supertranslations $x(\theta)$, which are two arbitrary functions on $S^1$. The expansion along the integral curve with infinitesimal parameter $t$ is
\begin{equation}
    x_t (\theta) = t \alpha (\theta) + O (t^2), \qquad f_t (\theta) = \theta + t u (\theta) + O(t^2).
\end{equation}
The superrotation is actually the diffeomorphism of the $S^1$, and $\alpha, u,x$ are the vector fields on $S^1$, as discussed in section \ref{sec:Virasoro}.

The centrally extended BMS$_3$ algebra is given by
\begin{align*}
    & [(u, \lambda; \alpha, \mu), (v, \rho; \beta, \nu)] = ([u, v], -\<s(u), v\>_0; [u, \beta] - [v, \alpha], \<s(v), \alpha\>_0 - \<s(u), \beta\>_0),
\end{align*}
where $\<\cdot, \cdot \>_0$ is defined in \eqref{eq:innerVirac}. The inner product is defined as
\begin{equation}
    \<(j, c_1; p, c_2), (u, \lambda; \alpha, \mu)\> = \<j, u\>_0 + \<p, \alpha\>_0 + c_1 \lambda + c_2 \mu.
\end{equation}
Here, the supermomentum $p$ and angular supermomentum $j$ are coadjoint vector dual to the Witt algebra, and transform as \eqref{eq:Adf*b}. The finite coadjoint action is \cite{Oblak:2016eij}
\begin{equation} \label{eq:finiteBMS3}
    \begin{split}
        \widehat{\text{Ad}}^*_{(f, x)} (j, c_1; p, c_2) = & \bigg(\text{Ad}^*_f j - c_1 S(f^{-1}) + \text{ad}^*_x \left[ \text{Ad}^*_f p - c_2 S(f^{-1}) \right] + c_2 s(x), c_1; \\
        & \text{Ad}^*_f p - c_2 S(f^{-1}), c_2\bigg).
    \end{split}
\end{equation}
The infinitesimal coadjoint action $\widehat{\text{ad}}^*_{(u, \alpha)} (j, c_1; p, c_2) = (\delta j, 0; \delta p, 0)$ is then derived as
\begin{equation} \label{eq:deltapj}
    \begin{split}
        & \delta p = - u (\theta) p' (\theta) - 2 u' (\theta) p (\theta) + \frac{c_2}{24 \pi} u'''(\theta), \\
        & \delta j = - u j' - 2 u' j  + \frac{c_1}{24 \pi} u''' - \alpha p' - 2 \alpha' p + \frac{c_2}{24 \pi} \alpha'''.
    \end{split}
\end{equation}
The BMS$_3$ algebra can be reproduced by the Poisson bracket of the coadjoint basis $j_n, p_n$, i.e., the coefficients in the Fourier expansion of $j, p$.

Perturbatively, the geometric action is derived as
\begin{equation} \label{eq:BMS3A}
    \begin{split}
        A & = - \left\< \widehat{\text{Ad}}^*_{(f, x)} (j_0, c_1; p_0, c_2), (u, m_u; \alpha, m_\alpha) \right\> \\
        & \approx 2 \pi i \sum_{n \in \mathbb Z} \int_{-\beta}^0 \td t (\dot F_n \mathcal{A}_n + 2 \dot x_n \mathcal{B}_n) n F_{-n}, 
    \end{split}
\end{equation}
where the MC-from is given by
\begin{equation}
    \begin{split}
        (u, m_u; \alpha, m_\alpha) & = \delta_{(f, x)} [(f, \lambda; x, \mu) (g, \rho; y, \nu)] |_{(g, \rho; y, \nu)= (f, \lambda; x, \mu)^{-1}} \\
        & = (u, m_u; \delta x - [u, x], \delta \mu + \<s(u), x\>_0).
    \end{split}
\end{equation}
Here, $(u, m_u)$ is the MC form of Virasoro \eqref{eq:muVir}.

The charges are spanned by the basis $L_k, M_k$ defined as
\begin{align} 
        L_k & = \left\< \widehat{\text{Ad}}^*_{(f, x)} (j_0, c_1; p_0, c_2), (e^{-ik \theta}, 0; 0,0) \right\> \nonumber\\ & \approx
        \begin{cases}
            & 2 \pi j_0 + 4 \pi \sum_{n =1}^\infty n^2 \left[ F_n F_{-n} \mathcal{A}_n + (F_n x_{-n} + x_n F_{-n}) \mathcal{B}_n \right], ~~ k =0, \label{eq:BMS3Ln}\\
            & 4 \pi ik (F_{- k} \mathcal{A}_k - x_{- k} \mathcal{B}_k), \qquad k \neq 0,\\
        \end{cases}\\
        M_k & = \left\< \widehat{\text{Ad}}^*_{(f, x)} (j_0, c_1; p_0, c_2), (0, 0; e^{-ik \theta},0) \right\> \nonumber\\ & \approx 
        \begin{cases}
            & 2 \pi p_0 + 4 \pi \sum_{n =1}^\infty F_n F_{-n} n^2 \mathcal{B}_n, \quad k = 0,\label{eq:BMS3Mn} \\ 
            & 4 \pi ik F_{- k} \mathcal{B}_k, \qquad k \neq 0.
        \end{cases}
\end{align}
The Poisson bracket \eqref{eq:PoissonPhi} gives the BMS$_3$ algebra
\begin{equation} \label{eq:cBMS3A}
    \begin{split}
        [L_n, L_m] & = i\{L_n, L_m\} =  (n-m) L_{n+m} + \frac{c_1}{12} n^3 \delta_{n+m,0}, \\
        [L_n, M_m] & = i \{L_n, M_m\} = (n-m) M_{n+m} + \frac{c_2}{12} n^3 \delta_{n+m,0}, \\
        [M_n, M_m] & = i \{M_n, M_m\} = 0.
    \end{split}
\end{equation}

\subsection{Classifying the 3D flat boundary graviton} \label{sec:bdygravitonbms3}

The solution space of 3D asymptotically flat gravity with prescribed boundary conditions at null infinity was derived in \cite{Barnich:2010eb}. The line-element is given by
\begin{equation} \label{eq:flat3d}
    \td s^2 = \Theta(\phi) \td u^2 - 2 \td u \td r + (\Xi (\phi) + u \Theta'(\phi)) \td u \td \phi + r^2 \td \phi^2.
\end{equation}
The finite or infinitesimal BMS$_3$ transformation at null infinity yields the transformation of $\Theta, \Xi$, which is the same as \eqref{eq:finiteBMS3} or \eqref{eq:deltapj} with the identification \cite{Barnich:2013yka, Barnich:2012rz}
\begin{equation} \label{eq:j0p0XiTh}
    (j_0, p_0) = \frac{1}{16 \pi G} (\Xi, \Theta), \qquad c_1 = 0, \quad c_2 = \frac{3}{G}.
\end{equation}
Therefore, for the same reason as in the AdS$_3$ case, the vacuum solution \eqref{eq:flat3d} can be classified by the coadjoint orbits $W_{(j_0,c_1; p_0,c_2)}$ of BMS$_3$ group \cite{Barnich:2015uva, Barnich:2014zoa}, which is labeled by reference points $(j_0, p_0)$ determined by the stabilizer $\delta p_0 = \delta j_0 = 0$. Note that $p_0$ shares the same equation of the stabilizer of Virasoro group. The BMS$_3$ orbit $W_{(j_0,c_1; p_0,c_2)}$ is then a bundle over the cotangent bundle of the Virasoro orbit $\mathcal{W}_{(p_0, c_2)}$, with the stabilizer of $p_0$ being the little group of BMS$_3$ \cite{Barnich:2015uva}. 

Let us first see the constant orbits determined by $\delta p_0 = \delta j_0 = 0$ with $p'_0 = j'_0 = 0$ in \eqref{eq:deltapj}, the solution of which can be classified as
\begin{equation} \label{eq:stabjp}
    \mathfrak{stab} (j_0, p_0)=
    \begin{cases}
        & \mathcal{B}_n \neq 0 : u = u_0; \quad \alpha = \alpha_0, \\
        & \mathcal{B}_n = \mathcal{A}_n = 0: u = u_0, e^{\pm in \theta}; \quad \alpha = \alpha_0, e^{\pm in \theta} \\
        & \mathcal{B}_n = 0, ~~ \mathcal{A}_n \neq 0: u = u_0; \quad \alpha = \alpha_0, e^{\pm in \theta}
    \end{cases}
\end{equation}
where $u_0, \alpha_0$ are constants, and the coefficients $\mathcal{A}_n, \mathcal{B}_n$ are given by
\begin{equation}
    \mathcal{A}_n = j_0 + \frac{c_1 n^2}{48 \pi}, \qquad \mathcal{B}_n = p_0 + \frac{c_2 n^2}{48 \pi}.
\end{equation}

Then, following the approach in \cite{Witten:1987ty}, we perturbatively discuss the non-constant orbits $(j_0, p_0)$ of BMS$_3$ group near the constant orbits $(P_0, J_0)$
\begin{equation}
    p_0 = P_0 + \sum_{n \in \mathbb Z, n\neq 0} P_n e^{in \theta}, \qquad j_0 = J_0 + \sum_{n \in \mathbb Z, n\neq 0} J_n e^{in \theta},
\end{equation}
with $(P_n, J_n) \ll (P_0, J_0)$. If $(p_0, j_0)$ is on the orbit of $(P_0, J_0)$, then there exists an infinitesimal transformation \eqref{eq:deltapj} that covers all modes $(P_n, J_n)$, namely the following equations have solutions for $(u_n, \alpha_n)$ for any $n$
\begin{equation}
    \frac{i}{2} J_n = u_n \left( J_0 + \frac{n^2 c_1}{48 \pi} \right) + \alpha_n \left( P_0 + \frac{n^2 c_2}{48 \pi} \right), \quad \frac{i}{2} P_n = u_n \left( P_0 + \frac{n^2 c_2}{48 \pi} \right).
\end{equation}
For $P_0 \neq - \frac{c_2 k^2}{48 \pi} (\forall k \in \mathbb Z_+)$, the solution for any $n \in \mathbb Z$ always exists
\begin{equation}
    u_n = \frac{i P_n/2}{P_0 + \frac{n^2 c_2}{48 \pi}}, \quad \alpha_n = \frac{i J_n/2}{P_0 + \frac{n^2 c_2}{48 \pi}} - \frac{i}{2} P_n \frac{J_0 + \frac{n^2 c_1}{48 \pi}}{(P_0 + \frac{n^2 c_2}{48 \pi})^2}.
\end{equation}
For $P_0 = - \frac{c_2 k^2}{48 \pi} (k \in \mathbb Z_+)$, the modes $(J_n, P_n),~ n \neq \pm k$ can be set to zero with $(u_n, \alpha_n), |n| \neq k$ given above, while $P_{\pm k}$ cannot. Then, if $J_0 \neq - \frac{c_1 k^2}{48 \pi}$, the $J_{\pm k}$ modes can be further set to zero by
\begin{equation}
    u_{\pm k} = \frac{i J_{\pm k}/2}{J_0 + \frac{k^2 c_1}{48 \pi}}, \qquad \forall \alpha_{\pm k}.
\end{equation}
However, $J_{\pm k}$ cannot be set to zero if $J_0 = - \frac{c_1 k^2}{48 \pi}$. As a result, there are two types non-constant orbits at the perturbative level
\begin{equation} \label{eq:nonconstbms3}
    p_0 = - \frac{c_2 k^2}{48 \pi} + P_k e^{i k \theta} + P_{-k} e^{-i k \theta}, \quad j_0 = 
    \begin{cases}
        & \text{Constant} \neq \frac{c_1 k^2}{48 \pi} \\
        & - \frac{c_1 k^2}{48 \pi} + J_k e^{i k \theta} + J_{-k} e^{-i k \theta}
    \end{cases}.
\end{equation}
However, a consistent analysis of the non-perturbative solutions for BMS$_3$ non-constant orbits can be a fiendishly challenging
task, unlike the Virasoro case. We therefore restrict our analysis in this paper to constant orbits.

\begin{figure}[htbp]
    \centering
    \includegraphics[width=0.5\linewidth]{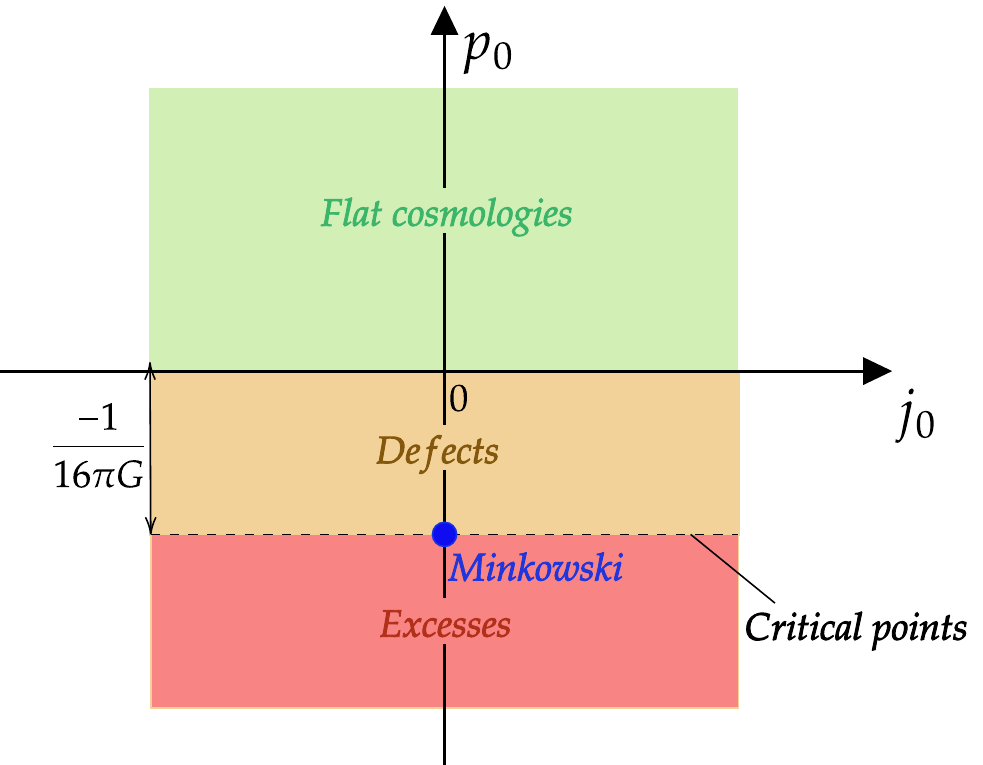}
    \caption{Asymptotic flat phase space with constant mass and angular momentum.}
    \label{fig:flatphase}
\end{figure}

The phase space of the familiar solutions \cite{Barnich:2012xq, Bagchi:2012xr, Deser:1983tn} (see also \cite{Compere:2018aar}) correspond to constant orbits and are divided through the values of $(j_0, p_0)$, as depicted in Fig. \ref{fig:flatphase}, which is adapted from the figures in \cite{Barnich:2012aw, Compere:2018aar}. Here, we present the isometry (stabilizer) of each region:
\begin{enumerate}
    \item \textbf{Flat cosmology:} $p_0>0$, its isometry generated by $\{L_0; M_0\}$.
    \item \textbf{Conical defects:} $-\frac{c_2}{48 \pi} < p_0 <0$, its isometry is generated by $\{L_0, M_0\}$.
    \item \textbf{Conical excesses:} $p_0 < - \frac{c_2}{48 \pi}$, its isometry is generated by
    \begin{enumerate}
        \item $\{L_0; M_0\}$ for $p_0 \neq -\frac{c_2 k^2}{48 \pi} (\forall k \in \mathbb Z_+)$
        \item $\{L_0, L_{\pm k}; M_0, M_{\pm k}\}$ for $p_0 = -\frac{c_2 k^2}{48 \pi} (k \in \mathbb Z_+), ~ j_0 = 0$.
        \item $\{L_0; M_0, M_{\pm k}\}$ for $p_0 = -\frac{c_2 k^2}{48 \pi} (k \in \mathbb Z_+), ~ j_0 \neq 0$.
    \end{enumerate}
    \item \textbf{Critical points between defects and excesses:} $p_0 = -\frac{c_2}{48 \pi}$. When $j_0 \neq 0$, the isometry is generated by $\{L_0; M_0, M_{\pm 1}\}$. When $j_0 = 0$, the solution is \textit{Global Minkowski} and the generators of its isometry are $\{L_0, L_{\pm 1}; M_0, M_{\pm 1}\}$, which is 3D Poincar\'e algebra.
\end{enumerate}

\subsection{Coadjoint modules for constant orbits}

Now we investigate coadjoint modules for constant BMS$_3$ orbits with the Hamiltonian $\tilde H = M_0 - i \frac{\xi}{\beta} L_0$ through path integral quantization at one-loop level. We should first express the geometric action and the generators $L_0, M_0$ by Darboux coordinates. At leading order, using the approximation $F\approx\theta$ and $x\approx0$, the KK form on the constant orbit is given by
\begin{equation} \label{eq:OmegaBMS3}
    \begin{split}
        \Omega_b \left( \widehat{\text{ad}}^*_{(u, \alpha)} b, \widehat{\text{ad}}^*_{(v, \beta)} b \right) & = \left\< \widehat{\text{Ad}}^*_{(f, x)} (j_0, c_1; p_0, c_2), [(u, 0; \alpha, 0), (v, 0; \beta, 0)] \right\> \\
        & \approx 4 \pi i \sum_{m, n \in \mathbb Z} \delta_{m+n} m [u_m v_n \mathcal{A}_m + (u_m \beta_n - v_m \alpha_n) \mathcal{B}_m].
    \end{split}
\end{equation}
The symplectic matrix $\Omega_b$ is diagonalized in blocks $\omega_n$ for each $n \neq 0$. The symplectic matrix $\omega_n$ admits three distinct cases, precisely corresponding to the three classes of stabilizers in \eqref{eq:stabjp}. In each case, the ranges of $\mathcal{A}_n, \mathcal{B}_n$ coincide with those characterizing the corresponding stabilizer. We will analyze the Darboux coordinates, coadjoint modules, and characters for each class in turn.

\subsubsection{The stabilizer with $\mathcal{B}_n \neq 0~ (\forall n \in \mathbb Z_+)$} \label{sec:Bnneq0}

For generic supermomentum ($\mathcal{B}_n \neq 0$), the stabilizer \eqref{eq:stabjp} is
\begin{equation}
    \mathfrak{stab} (j_0, p_0) = \{ (1, 0), (0, 1) \} = \{L_0, M_0\}.
\end{equation}
In this case, the block $\omega_n$ is
\begin{equation}
    \omega_n = - \omega_{-n} = 4 \pi i n
    \begin{pmatrix}
        \mathcal{A}_n & \mathcal{B}_n \\
        -\mathcal{B}_n & 0
    \end{pmatrix}.
\end{equation}
Therefore, the non-vanishing Poisson brackets for coordinates in the coadjoint orbit are derived from the inverse of $\omega_n$ as 
\begin{equation}
    \{x_{-n}, x_n\} = (\omega_{-n}^{-1})_{22} = \frac{i \mathcal{A}_n}{4 \pi n \mathcal{B}^2_n}, \qquad \{F_{-n}, x_n\} = (\omega_{-n}^{-1})_{12} = -\frac{i}{4 \pi n \mathcal{B}_n},
\end{equation}
where $x_n (t)$ is the coefficient of the Fourier expansion for the supertranslation parameter $x_t = \sum_n e^{-i n \theta} x_n (t)$, and $F_n$ is defined in \eqref{eq:Fexpansion}. The reality condition for $x$ also yields the constraint for its coefficient: $x_{-n} = x_n^*$. There are two sets of Darboux coordinates \cite{Garbarz:2015lua}
\begin{equation} \label{eq:DarbouxBMS3}
    \begin{split}
        q^x_n & = - \sqrt{2 \pi \mathcal{B}_n} i (F_n - F_{-n}), \quad q^y_n = -\sqrt{2 \pi \mathcal{B}_n} (F_n + F_{-n}), \\
        p^x_n & = n \sqrt{2 \pi \mathcal{B}_n} \left[ x_n + x_{-n} + \frac{\mathcal{A}_n}{2 \mathcal{B}_n} (F_n + F_{-n}) \right], \\
        p^y_n & = i n \sqrt{2 \pi \mathcal{B}_n} \left[ x_{-n} - x_n + \frac{\mathcal{A}_n}{2 \mathcal{B}_n} (F_{-n} - F_n) \right],
    \end{split}
\end{equation}
for any $n>0$, and the non-vanishing Poisson brackets are
\begin{equation}
    \{q^x_n, p^x_n\} = \{q^y_n, p^y_n\} = 1.
\end{equation}

In this case, the geometric action, and the charges $L_0, M_0$ are reformulated by Darboux coordinates as
\begin{equation} \label{eq:ALM}
    \begin{split}
        A & = -\int_{-\beta}^0 \td t \sum_{n \in \mathbb Z_+} [p_n^x \dot q_n^x + p_n^y \dot q_n^y], \\
        M_0 & = 2 \pi p_0 + \sum_{n \in \mathbb Z_+} \frac{n^2}{2} (q_n^r)^2, \quad (q_n^r)^2 = (q_n^x)^2 + (q_n^y)^2, \\
        L_0 & = 2 \pi j_0 + \sum_{n \in \mathbb Z_+} n J_n, \quad J_n = q_n^x p_n^y - q_n^y p_n^x,
    \end{split}
\end{equation}
where $J_n$ is the angular momentum on the $(q_n^x, q_n^y)$ plane and $e^{i J_n n \xi}$ generates the rotation by angle $n \xi$.

The contour of a converged definition of path integral in this case is given by
\begin{equation}
    x_n^* = \begin{cases}
        x_{-n}, & n > n_0 \\ -x_{-n}, &1 \leq n \leq n_0 
    \end{cases}, \qquad F_n^* = \begin{cases}
        F_{-n}, & n > n_0 \\ -F_{-n}, &1 \leq n \leq n_0
    \end{cases}.
\end{equation}
By contrast, for $\mathcal{B}_1>1$, the Hamiltonian has a lower bound. This case describes the conical defect solutions and the flat cosmologies in Fig. \ref{fig:flatphase}. The contour is defined as
\begin{equation}
    x_n^* = x_n, \qquad F_n^* = F_{-n}, \qquad \forall n \in \mathbb Z_+.
\end{equation}
The transition amplitude on these orbits is therefore given by the path integral as
\begin{equation}
\bra{q^{(N)}} e^{- \hat M_0 \beta + i \xi \hat L_0} \ket{q^{(0)}}  = \prod_{n \in \mathbb Z_+} \bra{q^{(N)}_n} e^{- \hat M_{0,n} \beta + i \xi \hat L_{0, n}} \ket{q^{(0)}_n}, \label{eq:BMSPIBkneq0}     
\end{equation}
where
\begin{multline}
    \bra{q^{(N)}_n} e^{- \hat M_{0,n} \beta + i \xi \hat L_{0,n}} \ket{q^{(0)}_n} \\
    = \int^{q_n (0) = q^{(N)}_n}_{q_n (-\beta) = q^{(0)}_n} D^2q_n \int D^2p_n e^{- \int^0_{-\beta} \td t [ip_n^x \dot q_n^x + ip_n^y \dot q_n^y - i \frac{\xi}{\beta} n J_n + \frac{n^2}{2} (q_n^r)^2 ]}.
\end{multline}
Here, we introduce the two-dimensional vectors of the Darboux coordinates $p_n= (p_n^x ~~ p_n^y)^T,~ q_n= (q_n^x ~~ q_n^y)^T$ and their measures $D^2 p_n = Dp_n^x Dp_n^y, D^2q_n = Dq_n^x Dq_n^y$ for notational brevity, where $T$ means transverse operation to the matrix. Moreover, there is an  abbreviation $q^{(k)} = (q^{(k)}_1, \cdots, q^{(k)}_n)$ with the superscripts $(k) = (0), (N)$ denoting the value of $q$ at $t=- \beta$ and $t=0$, respectively. We can introduce a similar notation for $p^{(k)}$, as will be discussed later. The path integral is defined by the contour living in the complex planes of $x_n$ and $F_n$. When $\mathcal{B}_{n_0}<0<\mathcal{B}_{n_0+1} (n_0 \in \mathbb Z_+)$, the Hamiltonian is unbounded from below. These orbits belong to the conical excesses region in Fig. \ref{fig:flatphase}.

The path integral in \eqref{eq:BMSPIBkneq0} can be directly calculated by discretization
\begin{equation}
\begin{split}
    &\bra{q^{(N)}_n} e^{- \hat M_{0,n} \beta + i \xi \hat L_{0,n}} \ket{q^{(0)}_n} \\
    =& \int \prod_{i=1}^{N-1} D^2q_n^{(i)} \prod_{j=0}^{N-1} D^2p_n^{(j)} e^{- i(p_n^{(j)})^T (q_n^{(j+1)} - R (\frac{n\xi}{N}) q_n^{(j)}) - \frac{n^2}{2} (q_n^{(j)})^T q_n^{(j)} \frac{\beta}{N}} \\
    = &\int \prod_{i=1}^{N-1} D^2q_n^{(i)} \prod_{j=0}^{N-1} \delta^2 \left(q_n^{(j+1)} - R \left(\frac{n\xi}{N} \right) q_n^{(j)} \right) e^{- \frac{n^2}{2} (q_n^{(j)})^T q_n^{(j)} \frac{\beta}{N}}, \label{eq:BMS3amplituden}
\end{split}
\end{equation}
where the matrix $R$ is given by
\begin{equation}
    R \left(\frac{n\xi}{N} \right) = I + \frac{B}{N}, \qquad B = \begin{pmatrix}
        0 & -n\xi \\ n\xi & 0
    \end{pmatrix},
\end{equation}
which generates a rotation by the infinitesimal angle $\frac{n\xi \beta}{N}$ at the continuous limit $N \to \infty$, as one can easily check that
\begin{equation}
\begin{split}
    \lim_{N \to \infty} \left[ R \left(\frac{n\xi}{N} \right) \right]^N & = \lim_{N \to \infty} \left(I+\frac{B}{N} \right)^N \\
    &= e^B
    = \begin{pmatrix}
        \cos (n \xi) & - \sin (n \xi) \\ \sin (n \xi) & \cos (n \xi)
    \end{pmatrix} = R(n\xi).
    \end{split}
\end{equation}
The delta functions in \eqref{eq:BMS3amplituden} gives the following constraints
\begin{equation}
    q_n^{(j)} = R \left(\frac{n\xi j}{N} \right) q_n^{(0)}, \qquad \forall j = 0, 1, \cdots, N,
\end{equation}
indicating that $(q_n^{(j)})^T q_n^{(j)}$ is constant for any $j$, as the rotation matrix satisfies $R^T R=I$. From the continuous point of view, it means that $q_n^T q_n = A_n^2$, with $A_n$ being a real constant independent of $t$. As a result, \eqref{eq:BMS3amplituden} can be simply derived as
\begin{equation} \label{eq:Poincareamplituden}
    \bra{q^{(N)}_n} e^{- \hat M_{0,n} \beta + i \xi \hat L_{0,n}} \ket{q^{(0)}_n} = \delta^2 \left(q_n^{(N)} - R (n\xi) q_n^{(0)} \right) e^{- \frac{n^2}{2} A_n^2 \beta},
\end{equation}
where the delta function comes from the norm $\< q_n^{(N)} \ket{R (n\xi) q_n^{(0)}}$.

The reference state $\ket{j_0, p_0}$ within the BMS$_3$ coadjoint module is prepared by taking $\beta \to \infty$ in \eqref{eq:BMSPIBkneq0}. For $A_n \neq 0$, the path integral is exponentially suppressed by the Boltzmann factor $e^{- \frac{n^2}{2} A_n^2 \beta}$. Consequently, the path integral is dominated by configurations with $A_n = 0$, implying that $q^x_n = q^y_n = 0$, which is equivalent to the following rest-frame condition based on the definition of Darboux coordinates in \eqref{eq:DarbouxBMS3}
\begin{equation} \label{eq:annihilateM}
    F_n \ket{j_0, p_0} = M_n \ket{j_0, p_0} = 0, \qquad \forall n \neq 0.
\end{equation}
One can easily check that the regularity condition \eqref{eq:regular} in this case is automatically satisfied. Moreover, the resulting BMS module should not include the modes $L_0, M_0$, as they are the generators within the stabilizer. Their precise action can be determined from \eqref{eq:BMS3Ln}, \eqref{eq:BMS3Mn}, after imposing the normal ordering $:\cdots:$, which places $F_n$ to the right of $x_n$ and orders the $x_n$ modes such that modes with smaller indices appear to the left. There is no need to ordering the $F_n$ modes here, since they commute with each other. When acting on the reference state, all subleading corrections to $M_0, L_0$ in \eqref{eq:BMS3Ln} and \eqref{eq:BMS3Mn} vanish, since every such correction contains at least one mode $F_{n\neq 0}$, which is moved by the normal-ordering prescription to the right and hence annihilates the reference state. We therefore find
\begin{equation} \label{eq:eigenLM}
    L_0 \ket{j_0, p_0} = 2 \pi j_0 \ket{j_0, p_0}, \qquad M_0 \ket{j_0, p_0} = 2\pi p_0 \ket{j_0, p_0}.
\end{equation}
Actually, here $\ket{j_0, p_0}$ is the common eigenstate of $L_0, M_0$, which requires that they commute with each other. Fortunately, based on the BMS$_3$ algebra \eqref{eq:cBMS3A}, we can easily observe that $[M_0, L_0] = 0$. The resulting coadjoint module is then spanned by
\begin{equation} \label{eq:Bnneq0module}
    :L_{n_i} \cdots L_{n_1}: \ket{j_0, p_0}, \qquad \forall n_1, \cdots, n_i \neq 0,
\end{equation}
where normal ordering places $L_n$ with larger indices on the right. 

From the BMS$_3$ algebra \eqref{eq:cBMS3A}, we find that the state with the above form is also the eigenstate of $L_0$ with the eigenvalue being $2 \pi j_0 - (n_1 + \cdots + n_i)$. Moreover, $M_0$ is an upper triangular matrix with all diagonal elements being $2 \pi p_0$. Consequently, the character can be calculated by counting the states in the module and using the Euler function \eqref{eq:Euler} as
\begin{equation} \label{eq:BMSchara1}
    \begin{split}
        \text{Tr} e^{- M_0 \beta + i \xi L_0} & = \sum_{M,N=0}^\infty e^{i \xi (2 \pi j_0 + M - N) - 2 \pi \beta p_0} p(M) p(N) \\
        & = \frac{e^{2 \pi (i \xi j_0 - \beta p_0)}}{\varphi (e^{i \xi}) \varphi (e^{-i \xi})} = \frac{e^{2 \pi (i \xi j_0 - \beta p_0)}}{\prod_{n =1}^\infty |1-e^{in \xi}|^2}.
    \end{split}
\end{equation}
There is a subtlety when $\xi/2\pi$ is a rational number. Namely, for $\xi=2\pi\frac{a}{b}$ with $a, b$ coprime integers, the resonance modes with $n = mb (\forall m \in \mathbb Z_+)$ give divergent contributions to the expression above, as for these modes we always have $e^{i mb \xi} = e^{2 \pi i m a} = 1$. In the following, we will show from the path-integral perspective that these are infrared (IR) divergences proportional to a collection of $\delta^2(0)$ factors.

The character can be reproduced by path integral \eqref{eq:chiHEucl}, which gives a collection of the path integral of 3D Poincar\'e single-particle
\begin{equation}
    \text{Tr} e^{- M_0 \beta + i \xi L_0} = e^{2 \pi (i \xi j_0 - \beta p_0)} \prod_{n =1}^\infty \chi^P_{(n \xi)},
\end{equation}
where $\chi^P_{(n \xi)}$ of 3D Poincar\'e group can be derived by transforming to the momentum basis as in \cite{Garbarz:2015lua}, or by directly calculation of path integral using \eqref{eq:Poincareamplituden}
\begin{equation} \label{eq:3dPoncarePI}
    \begin{split}
        \chi^P_{(n\xi)} & = \int D^2q^{(0)}_n D^2q^{(N)}_n \delta (q^{(N)}_n -q^{(0)}_n) \bra{q^{(N)}_n} e^{- \hat M_{0,n} \beta + i \xi \hat L_{0,n}} \ket{q^{(0)}_n} \\
        & = \int D^2q^{(0)}_n \delta^2 \left[(I - R (n\xi)) q_n^{(0)} \right] e^{- \frac{n^2}{2} (q^{(0)}_n)^T q^{(0)}_n \beta}.
    \end{split}
\end{equation}
When $\xi/ (2\pi)$ is an irrational number, the matrix $I - R(n \xi) \neq 0$. In this case, the result of \eqref{eq:BMSchara1} will be simply reproduced by using
\begin{equation}
    \delta^2 \left[(I - R (n\xi)) q_n^{(0)} \right] = \frac{\delta^2 (q_n^{(0)})}{\mathrm{det} [I - R (n\xi)]}, \quad \mathrm{det} [I - R (n\xi)] = |1 - e^{i n \xi}|^2.
\end{equation}
However, when $\xi/(2 \pi) = a/b$ is rational, then $R(\xi bm) = I$. In this case, we have
\begin{equation}
    \chi^P_{(\xi mb)} = \delta^2 (0) \int D^2q^{(0)}_{mb} e^{- \frac{(mb)^2}{2} (q^{(0)}_{mb})^T q^{(0)}_{mb} \beta} = \delta^2 (0) \frac{2\pi}{m^2 b^2 \beta}.
\end{equation}
The character is then
\begin{equation} \label{eq:IRBMS3general}
    \text{Tr} e^{- M_0 \beta + i \xi L_0} = \frac{e^{2 \pi (2 \pi i j_0 a/b - \beta p_0)}}{\prod_{n \in \mathbb Z_+, n \notin \{mb| \forall m \in \mathbb Z_+\}} |1-e^{in \xi}|^2} \prod_{m \in \mathbb Z_+} \left(\delta^2 (0) \frac{2\pi}{m^2 b^2 \beta} \right),
\end{equation}
where the divergence originates from the norm $\<q\ket{q} = \delta^2 (0)$ of the continuous eigenbasis $\ket{q}$, which is equivalent to the infinite volume in the phase space. The resulting $\delta^2(0)$ divergence is therefore an infrared (IR) divergence. Since the divergent factor $\delta^2(0)$ factorizes from the result, we can simply define a finite character by dividing it. Similar issues arise in the recent study of the Minkowski spacetime partition function in \cite{Cotler:2024cia}, as well as in the analysis of magnetic Carrollian theories in \cite{Cotler:2024xhb}.

The little group of $\mathcal{B}_n \neq 0$ case is $U(1)$ with the generator $\{L_0\}$. From \eqref{eq:eigenLM} we can conclude that the representation \eqref{eq:Bnneq0module} is induced from the 1D little-group representation $R (\theta) = e^{i 2\pi j_0 \theta}$. Note that the massive representation with the positive mass $M = 2 \pi \mathcal{B}_1 = 2\pi p_0+\frac{c_2}{24}>0$ has been discussed in \cite{Barnich:2014kra, Campoleoni:2016vsh}. This massive regime is contained in our case $\mathcal{B}_n \neq 0 (\forall n \in \mathbb Z_+)$, and the corresponding reference state \cite{Campoleoni:2016vsh} and character \cite{Oblak:2015sea} agree with those in \eqref{eq:annihilateM}, \eqref{eq:eigenLM}, and \eqref{eq:BMSchara1}. 

Finally, we will show that the BMS module discussed here can be reproduced as a limiting case from two copies of Virasoro algebra $\{\mathcal{L}_n, c\} \oplus \{ \bar{\mathcal{L}}_n, \bar c\}$, with $\mathcal{L}_n, \bar{\mathcal{L}}_n$ separately satisfying \eqref{eq:VirasorA} with central charges $c, \bar c$, respectively. Firstly, one can explicitly reproduce the BMS algebra \eqref{eq:cBMS3A} from the Virasoro algebras through the ultra-relativistic (UR) limit $\epsilon \to 0$
\begin{equation} \label{eq:ultraR}
    (L_n, c_1) = (\mathcal{L}_n + \bar{\mathcal{L}}_{-n}, c- \bar c), \qquad (M_n, c_2) = \epsilon (\mathcal{L}_n - \bar{\mathcal{L}}_{-n}, c + \bar c).
\end{equation}
Consider the $u(1) \oplus \overline{u(1)}$ stabilizer of $\{\mathcal{L}_n, c\} \oplus \{ \bar{\mathcal{L}}_n, \bar c\}$, i.e., the reference state $\ket{b_0, \bar b_0}$ satisfies
\begin{equation}
    \begin{split}
        & \mathcal{L}_0 \ket{b_0, \bar b_0} = 2 \pi b_0 \ket{b_0, \bar b_0}, \qquad \mathcal{L}_n \ket{b_0, \bar b_0} = 0 \quad (\forall n \in \mathbb Z_+), \\
        & \bar{\mathcal{L}}_0 \ket{b_0, \bar b_0} = 2 \pi \bar b_0 \ket{b_0, \bar b_0}, \qquad \bar{\mathcal{L}}_n \ket{b_0, \bar b_0} = 0 \quad (\forall n \in \mathbb Z_+).
    \end{split}
\end{equation}
Note that the overlines are introduced merely to distinguish the two copies of the Virasoro algebra. Using \eqref{eq:ultraR}, one obtains
\begin{equation}
    \begin{split}
        & L_0 \ket{j_0, p_0} = 2 \pi j_0 \ket{j_0, p_0}, \qquad M_0 \ket{j_0, p_0} = 2\pi p_0 \ket{j_0, p_0}, \\
        & (M_n + \epsilon L_n) \ket{j_0, p_0} = (M_{-n} - \epsilon L_{-n}) \ket{j_0, p_0} = 0, \qquad (\forall n \geq 1),
    \end{split}
\end{equation}
where $(j_0, p_0) = (b_0 + \bar b_0, \epsilon (b_0 - \bar b_0))$ and $\ket{b_0, \bar b_0} = \ket{j_0, p_0}$. The above annihilation condition recovers \eqref{eq:annihilateM} and \eqref{eq:eigenLM} by taking $\epsilon \to 0$.

\subsubsection{The stabilizer with $\mathcal{B}_k = 0=\mathcal{A}_k $} \label{sec:Bk=Ak=0}

When $\mathcal{B}_k = 0=\mathcal{A}_k ~ (p_0 = \frac{-c_2 k^2}{48 \pi}, ~ j_0 = \frac{-c_1 k^2}{48 \pi})$ with $k>0$, the stabilizer \eqref{eq:stabjp} is
\begin{equation} \label{eq:stabAnBn0}
    \mathfrak{stab} (j_0, p_0) = \{ (1, 0), (e^{\pm ik \theta}, 0), (0, 1), (0, e^{\pm ik \theta}) \} = \{M_0, M_{\pm k}, L_0, L_{\pm k}\}.
\end{equation}
In this case, the $\omega_k$ should be completely removed from the symplectic matrix. Then, $n$ in the Darboux coordinates \eqref{eq:DarbouxBMS3} cannot be $k$, and \eqref{eq:ALM} becomes
\begin{equation} \label{eq:ALMAnBn0}
    \begin{split}
        & A = -\int_{-\beta}^0 \td t \sum_{n \in \mathbb Z_+/\{k\}} [p_n^x \dot q_n^x + p_n^y \dot q_n^y], \\
        M_0 = 2 \pi p_0 + & \sum_{n \in \mathbb Z_+/\{k\}} \frac{n^2}{2} (q_n^r)^2, \quad L_0 = 2 \pi j_0 + \sum_{n \in \mathbb Z_+/\{k\}} n J_n.
    \end{split}
\end{equation}

Similar to \eqref{eq:BMSPIBkneq0}, the path integral of the transition amplitude on this orbit is a collection of 3D Poincar\'e single-particles \eqref{eq:Poincareamplituden} but without the $k$-th particle. The $k>1$ cases belong to the conical excesses in Fig. \ref{fig:flatphase}, with a Hamiltonian that is unbounded from below. In this case, the convergent path integral is defined on the contour satisfying
\begin{equation}
    x_n^* = \begin{cases}
        x_{-n}, & n > k \\ -x_{-n}, &1 \leq n <k
    \end{cases}, \qquad F_n^* = \begin{cases}
        F_{-n}, & n > k \\ -F_{-n}, &1 \leq n <k
    \end{cases}.
\end{equation}
On the other hand, $k=1$ case is the Minkowski solution as in Fig. \ref{fig:flatphase}, possessing a Hamiltonian with a lower bound. In this case, the integral contour is given by
\begin{equation}
    x_n^* = x_n, \qquad F_n^* = F_{-n}, \qquad n = 2, 3, 4, \cdots.
\end{equation}
As a result, the transition amplitude is
\begin{equation}
    \bra{q^{(N)}} e^{- \hat M_0 \beta + i \xi \hat L_0} \ket{q^{(0)}} = \prod_{n \in \mathbb Z_+/ \{k\}} \delta^2 \left(q_n^{(N)} - R (n\xi) q_n^{(0)} \right) e^{- \frac{n^2}{2} A_n^2 \beta}.
\end{equation}
In particular, the transition amplitude of $k=1$ case agrees with that of Minkowski space derived in \cite{Cotler:2024cia}.

Following the same calculation as in section \ref{sec:Bnneq0}, the half-line path integral with $\beta \to \infty$ similarly gives the following annihilation conditions
\begin{equation} \label{eq:annihiAk=Bk=0}
    F_n \ket{j_0, p_0} = M_n \ket{j_0, p_0} = 0, \qquad (\forall n \neq 0, \pm k).
\end{equation}
For the orbit with exceptional supermomentum ($\mathcal{B}_k = 0$), the little group is $SL^{(k)} (2, \mathbb R)$, with its algebra $\{L_0, L_{\pm k}\}$ being the subalgebra of the stabilizer \eqref{eq:stabAnBn0}. In this case, we need to expand the $M_{\pm k}$ in \eqref{eq:BMS3Mn} to the next order
\begin{equation} \label{eq:M+-k}
    M_{\pm k} = \frac{-c_2}{48 \pi} \sum_{n \in \mathbb Z} n (n \pm k) (n^2 \mp kn - k^2) F_n F_{-n \mp k}.
\end{equation}
Each term in this expression contains at least one $F_n$ with $n \neq 0, k$, which annihilates the reference state. Consequently, we have $M_{\pm k} \ket{j_0, p_0} =0$. Furthermore, when $\mathcal{A}_k$ also vanishes, we need to expand $L_{\pm k}$ in \eqref{eq:BMS3Ln} to the next leading order
\begin{equation}
    \begin{split}
        L_{\pm k} \approx & - \frac{c_1}{48 \pi} \sum_{n \in \mathbb Z_+} F_{-n} F_{-n \mp k} n (n \pm k) (n^2 \mp nk - k^2) \\
        & + \frac{c_2}{12 \pi} \sum_{n \in \mathbb Z_+} x_{-n-k} F_n n (2k-n) (n^2-k^2).
    \end{split}
\end{equation}
It is explicit that $L_{\pm k}$ also annihilates the reference state because of $F_n (n \neq 0, k)$. For the same reason as in \eqref{eq:eigenLM}, the reference state is also the common eigenstate of $L_0, M_0$. As a result, the complete definition of the reference state is
\begin{equation} \label{eq:Bn=An=0anni}
    \begin{split}
        & L_{\pm k} \ket{j_0, p_0} = 0, \qquad M_n \ket{j_0, p_0} = 0, ~~ (\forall n \neq 0), \\
        &  L_0 \ket{j_0, p_0} = 2 \pi j_0 \ket{j_0, p_0}, \qquad M_0 \ket{j_0, p_0} = 2\pi p_0 \ket{j_0, p_0}.
    \end{split}
\end{equation}
The resulting coadjoint module is then spanned by
\begin{equation} \label{eq:Bn=An=0module}
    :L_{n_1} \cdots L_{n_i}: \ket{j_0, p_0}; \qquad \forall n_1, \cdots, n_i \neq 0, \pm k.
\end{equation}
Note that the little group representation in this case is a trivial representation of $SL^{(k)} (2, \mathbb R)$, since the reference state is the eigenstate of $L_0, L_{\pm k}$. 

Using the state-counting procedure, the character can be derived by subtracting the contributions of descendants containing $L_{\pm k}$ modes from \eqref{eq:BMSchara1}. A direct computation shows that this is equivalent to multiplying \eqref{eq:BMSchara1} by the factor $1-e^{i \xi k} -e^{-i \xi k} + 1 = |1-e^{i \xi k}|^2$, namely
\begin{equation} \label{eq:BMSchara2}
    \text{Tr} e^{- M_0 \beta + i \xi L_0} = \frac{e^{2 \pi (i \xi j_0 - \beta p_0)}}{\prod_{n \in \mathbb Z_+/\{k\}} |1-e^{in \xi}|^2},
\end{equation}
or by path integral
\begin{equation}
    \text{Tr} e^{- M_0 \beta + i \xi L_0} = e^{2 \pi (i \xi j_0 - \beta p_0)} \prod_{n \in \mathbb Z_+/\{k\}} \chi^P_{(n \xi)}.
\end{equation}
IR divergences still exist when $\xi = 2 \pi a/b$ is rational. In this case, the $mb (m \in \mathbb Z_+)$ modes should be replaced by the $\delta^2 (0) \frac{2\pi}{m^2 b^2 \beta}$ as in \eqref{eq:IRBMS3general}. One can also deal with the IR divergence in a similar approach as in section \ref{sec:Bnneq0}.

When $k = 1$, the module \eqref{eq:Bn=An=0module} and character \eqref{eq:BMSchara2} reduce to those of the vacuum induced representation with vanishing angular momentum $J= 2 \pi j_0 + \frac{c_1}{24} = 2 \pi \mathcal{A}_1 = 0$ and mass $M= 2 \pi p_0 + \frac{c_2}{24} = 2 \pi \mathcal{B}_1 = 0$ in \cite{Barnich:2014kra, Campoleoni:2016vsh, Oblak:2015sea}. Moreover, following a similar approach in section \ref{sec:Bnneq0}, this BMS coadjoint module can also be derived from that of two copies of Virasoro group with the stabilizer $sl^{(k)} (2, \mathbb R) \oplus \overline{sl^{(k)} (2, \mathbb R)}$ through the UR limit \eqref{eq:ultraR}.

\subsubsection{The stabilizer with $\mathcal{B}_k = 0, ~ \mathcal{A}_k \neq 0$}

When $\mathcal{B}_k = 0, ~ \mathcal{A}_k \neq 0 ~(j_0 \neq -\frac{c_1 k^2}{48 \pi}, ~ p_0 = -\frac{c_2 k^2}{48 \pi})$, the stabilizer \eqref{eq:stabjp} is
\begin{equation} \label{eq:stabAneq0}
    \mathfrak{stab} (j_0, p_0) = \{ (1, 0), (0, 1), (0, e^{\pm ik \theta}) \} = \{M_0, M_{\pm k}, L_0\}.
\end{equation}
In this case, the $\alpha_{\pm k}, \beta_{\pm k}$ degrees of freedom in \eqref{eq:OmegaBMS3} should be removed from the BMS$_3$ phase space. Then, $\omega_{-k} = 4 \pi i k \mathcal{A}_k$, and only $F_k$ modes exist in the phase space. At $n=k$, the Poison bracket reduces to the Virasoro case as in \eqref{eq:FFPoisonVir},
\begin{equation}
    \{F_k, F_{-k}\} = (\omega_{-k})^{-1} = \frac{i}{4 \pi k \mathcal{A}_k}.
\end{equation}
Then one can introduce one set of Darboux as in \eqref{eq:DarVir}
\begin{equation} \label{eq:DarpqBMS}
    q_k= \sqrt{2 \pi \mathcal{A}_k} (F_k + F_{-k}), \qquad p_k = i k \sqrt{2 \pi \mathcal{A}_k} (F_k - F_{-k}),
\end{equation}
with $\{q_k, p_k\}=1$. For $n \neq k$ blocks, the Darboux coordinates are also \eqref{eq:DarbouxBMS3}.

In this case, we also have 
\begin{equation} \label{eq:ALMAnneq0}
    \begin{split}
        A & = -\int_{-\beta}^0 \td t \bigg[p_k \dot q_k + \sum_{n \in \mathbb Z_+/\{k\}} (p_n^x \dot q_n^x + p_n^y \dot q_n^y ) \bigg], \\
        M_0 = 2 \pi p_0 + & \sum_{n \in \mathbb Z_+/\{k\}} \frac{n^2}{2} (q_n^r)^2, \quad L_0 = 2 \pi j_0 + \frac{1}{2} (p_k^2 + k^2 q_k^2) + \sum_{n \in \mathbb Z_+/\{k\}} n J_n.
    \end{split}
\end{equation}
Therefore, the excitation on this orbit can be viewed as a collection of 3D Poincar\'e single-particles but with the $k$-th particle replaced by a 1D harmonic oscillator with frequency $\xi k/ \beta$ and mass $m=\beta/ \xi$. The convergent path integral in this case is defined by choosing the contour in complex planes of $x_n (n \neq k)$ and $F_n (n \in \mathbb Z_+)$. Note that the contour in complex planes of $x_n, F_n (n \neq k)$ are the same as those given in section \ref{sec:Bk=Ak=0}, and the contour in complex plane of $F_k$ is given by 
\begin{equation}
    F_k^* = \begin{cases}
        F_{-k}, & \mathcal{A}_k > 0 \\ -F_{-k}, & \mathcal{A}_k < 0
    \end{cases}.
\end{equation}
Then, the transition amplitude is derived by the path integral as
\begin{equation}
    \begin{split}
        \bra{q^{(N)}} e^{- \hat M_0 \beta + i \xi \hat L_0} \ket{q^{(0)}} = & \int Dp_k Dq_k e^{i \int^0_{-\beta} \td t (p_k \dot q_k - \xi \frac{p_k^2 + k^2 q_k^2}{2 \beta})} \\
        & \qquad \times \prod_{n \in \mathbb Z_+/ \{k\}} \bra{q^{(N)}_n} e^{- \hat M_{0,n} \beta + i \xi \hat L_{0, n}} \ket{q^{(0)}_n}.
    \end{split}
\end{equation}
Using the transition amplitudes of Poincar\'e \eqref{eq:Poincareamplituden} and harmonic oscillator, one obtains
\begin{equation} \label{eq:PIBk=0Akneq0}
    \begin{split}
        \bra{q^{(N)}} e^{- \hat M_0 \beta + i \xi \hat L_0} \ket{q^{(0)}} = & \sqrt{\frac{\beta/\xi}{2 \pi i \sin (\xi k)}} e^{\frac{ik}{2 \sin (\xi k)} [((q^{(N)}_k)^2+ (q^{(0)}_k)^2) \cos (\xi k) - 2 q^{(N)}_k q^{(0)}_k]} \\
        & \times \prod_{n \in \mathbb Z_+/ \{k\}} \delta^2 \left(q_n^{(N)} - R (n\xi) q_n^{(0)} \right) e^{- \frac{n^2}{2} A_n^2 \beta}. 
    \end{split}
\end{equation}
Particularly, when $k>1$, the orbits belong to the excesses in Fig. \ref{fig:flatphase}, whose Hamiltonian is unbounded from below. On the other hand, the $k=1$ case describes the critical points except for the Minkowski solution, as depicted by the dashed line in Fig. \ref{fig:flatphase}. As we will see later, this case possesses a Hamiltonian with a lower bound but is quite different from the standard massive and vacuum representations in previous studies \cite{Barnich:2014kra, Campoleoni:2016vsh}.

The regularity conditions originating from half-line path integral with $\beta \to \infty$ for this case give the same annihilation conditions as \eqref{eq:annihiAk=Bk=0}. Moreover, the path integral \eqref{eq:PIBk=0Akneq0} in this case contains a harmonic oscillator, which is independent with $\beta$ and is the contribution of $L_0$. Its path-integral quantization leads to an extra annihilation condition. From the definition of the Darboux coordinates in \eqref{eq:DarpqBMS}, we find that $F_{\pm k}$ correspond to the creation or annihilation operators of this oscillator. However, this harmonic-oscillator path integral does not arise from the real part of the Hamiltonian; instead, it appears through a phase factor and corresponds to a Lorentzian harmonic-oscillator path integral. It therefore does not admit a regularity condition analogous to that of the Euclidean half-line path integral. Consequently, both $F_{\pm k}$ can be chosen as the annihilation operator. In what follows, we consider the polarization in which $F_{-k}$ is chosen as the annihilation operator; the results for the other polarization are obtained by replacing $k$ with $-k$. 

Under the polarization we have chosen, the annihilation condition is given by
\begin{equation}
    F_n \ket{j_0, p_0} = 0, \qquad (\forall n \neq 0, k).
\end{equation}
Using $L_n, M_n$ given by \eqref{eq:BMS3Ln} and \eqref{eq:BMS3Mn}, these conditions can be reformulated as
\begin{equation}
    L_k \ket{j_0, p_0} = 0, \qquad M_n \ket{j_0, p_0} = 0, ~~ (\forall n \neq 0, \pm k).
\end{equation}
Note that the little group for this case is also $SL^{(k)} (2, \mathbb R)$. However, the generators $L_{\pm k}$ are not included in the stabilizer \eqref{eq:stabAneq0}, indicating that the representation corresponding to the orbits $\mathcal{B}_k =0,~ \mathcal{A}_k \neq 0$ is induced from a non-trivial little group representation of $SL^{(k)} (2, \mathbb R)$. Indeed, from \eqref{eq:BMS3Ln}, it is clear that $L_{-k} \ket{j_0, p_0} \neq 0$ for $\mathcal{B}_k = 0, \mathcal{A}_k \neq 0$. In particular, for $k=1$, the representation is induced from a non-trivial representation of the little group $SL(2,\mathbb{R})$. This provides a new class of representations on a coadjoint orbit with a Hamiltonian bounded from below, which differs significantly from the standard massive and vacuum representations, since the little group of the former is $U(1)$ and the latter is induced from the trivial representation of $SL(2,\mathbb{R})$. Moreover, from a similar analysis given near \eqref{eq:M+-k} and \eqref{eq:eigenLM}, one can show that the reference state is also the eigenstate of the modes $M_k, M_{- k}, M_0, L_0$ with the eigenvalues being $0, 0, 2 \pi p_0, 2\pi j_0$. In conclusion, the reference state is defined as
\begin{equation} \label{eq:annihiAnneq0}
    \begin{split}
        & L_k \ket{j_0, p_0} = 0, \qquad M_n \ket{j_0, p_0} = 0, ~~ (\forall n \neq 0), \\
        & L_0 \ket{j_0, p_0} = 2 \pi j_0 \ket{j_0, p_0}, \qquad M_0 \ket{j_0, p_0} = 2\pi p_0 \ket{j_0, p_0}.
    \end{split}
\end{equation}
The resulting coadjoint module is therefore spanned by
\begin{equation}
    :L_{n_i} \cdots L_{n_1}: \ket{j_0, p_0}; \qquad \forall n_1, \cdots, n_i \neq 0, k.
\end{equation}
Taking the UR limit \eqref{eq:ultraR}, the BMS module here can be derived from the two-copies Virasoro coadjoint module with the stabilizer $u(1) \oplus \overline{sl^{(k)} (2, \mathbb R)}$.

The characters can be calculated by subtracting the contributions of the descendants containing $L_k$ mode in \eqref{eq:BMSchara1}, which is equivalent to multiplying \eqref{eq:BMSchara1} by the factor $1 - e^{-i \xi k}$, namely
\begin{equation} \label{eq:BMSchara3}
    \mathrm{Tr} e^{- M_0 \beta + i \xi L_0} = \frac{e^{2 \pi i \xi j_0 + \beta \frac{c_2 k^2}{24}}}{(1-e^{ik \xi}) \prod_{n \in \mathbb Z_+/\{k\}} |1-e^{in \xi}|^2},
\end{equation}
or by using the periodic path integral of 1D harmonic oscillator \eqref{eq:1dHarOPI} and the 3D Poincar\'e particle \eqref{eq:3dPoncarePI} as
\begin{equation}
    \mathrm{Tr} e^{- M_0 \beta + i \xi L_0} = e^{2 \pi (i \xi j_0 - \beta p_0)} \chi^{H (\beta/\xi)}_{(\frac{\xi k}{\beta}, -i \beta)} \prod_{n \in \mathbb Z_+/\{k\}} \chi^P_{(n \xi)},
\end{equation}
where the zero-point energy of the harmonic oscillator are absorbed into $j_0$ by the renormalization of the orbit parameter $j_0$, which is consistent with the the normal-ordering procedure. Similarly, the IR divergences also exist in the case when $\xi = 2 \pi a/b$, and the mode $n= mb (m \in \mathbb Z_+, n \neq k)$ should be replaced by $\delta^2(0)$. Furthermore, when $k = mb$, the contribution of the harmonic oscillator also correspond to an IR divergence, and the divergent term $(1-e^{ik \xi})^{-1}$ is actually proportional to $\delta (0)$ in this case.

\section{Concluding remarks} \label{sec:conclusion}

In this paper, we develop a systematic approach to explicitly constructing the modules associated with the coadjoint orbits of the ASG. We show that, at one-loop order, a convergent path integral for the geometric action on a given orbit can always be defined by an appropriate choice of integration contour, regardless whether the Hamiltonian is bounded from below. The reference state is then defined by the Euclidean half-line path integral. Path-integral quantization determines which generators annihilate the reference state, while the remaining generators create independent excitations when acting on the reference state. Each coadjoint module is spanned by a reference state and its descendants. We then construct the coadjoint modules of the constant orbits of Virasoro and BMS$_3$ groups using this formalism.

The Virasoro constant coadjoint orbits are classified by the stabilizers $U(1)$ and $SL^{(k)} (2, \mathbb R)$. The resulting coadjoint modules for the former consist of the reference state $\ket{b_0 \neq \frac{-c n^2}{48 \pi}} (\forall n \in \mathbb Z_+)$, and its descendants, generated by $\mathcal{L}_{-n} (n \in \mathbb Z_+)$. These are the highest-weight modules generated from primary states. For the orbit with stabilizer $SL^{(k)} (2, \mathbb R)$, the modules are spanned by the reference states $\ket{b_0 = \frac{-c k^2}{48 \pi}}$ and their descendants obtained by acting with the generators $\mathcal{L}_{-n} (n \in \mathbb Z_+/ \{k\})$. These are also highest-weight modules. For the case of $k=1$, it corresponds to the module of vacuum HWR.

The BMS$_3$ coadjoint modules turn out to be induced modules generated from different little-group representations. For generic supermomentum, the reference state is a rest-frame state annihilated by the supertranslation generators $M_{n \neq 0}$. This directly reproduces the standard induced BMS$_3$ module with the little group $U(1)$ generated by $L_0$. For the exceptional supermomentum, the little group is enlarged to $SL^{(k)} (2, \mathbb R)$ and includes two additional superrotation generators $L_{\pm k}$. This gives rise to two distinct possibilities. If $SL^{(k)}(2,\mathbb R)$ is contained in the stabilizer, the reference state is annihilated by both $M_n (\forall n \neq 0)$ and $L_{\pm k}$, corresponding to the representation induced from the trivial representation of the little group $SL^{(k)}(2,\mathbb R)$. Otherwise, $L_{\pm k}$ survives as the additional physical DOF appearing in the path integral as the harmonic oscillator with frequency $k$. This case corresponds to a module induced from a non-trivial representation of the little group $SL^{(k)}(2,\mathbb R)$. We further show that the BMS$_3$ results can be obtained from the two copies of Virasoro via the ultra-relativistic limit.

We also compute the characters of all constant Virasoro and BMS$_3$ coadjoint orbits by two different methods, namely, by directly counting the number of states within the modules and by the path integral of the geometric action with a periodic Euclidean direction. The agreement between the two approaches provides a consistency check of our formalism. Similar to the discussion in \cite{Cotler:2024cia}, in the BMS$_3$ case, when $\xi/(2\pi)$ is rational, the trace structure of the character leads to infrared divergences due to our choice of an eigenbasis with continuous spectrum. The IR divergence indicates a subtlety in the choice of vacuum for the path integral with periodic boundary conditions. Similar subtleties arise in the selection of Carrollian vacuum, see, e.g., in \cite{Cotler:2024xhb, Fredenhagen:2026pia}. Fortunately, these IR divergences do not affect the half-line path integral used to prepare the reference state, since the associated delta-function factors are independent of $\beta$ and the $\delta(0)$ IR divergence arises only upon taking the trace. The resulting modules defined by the half-line path integral are therefore unaffected. This further underscores the importance of the module and transition amplitudes (or S-matrix): as emphasized in the introduction, for infinite-dimensional groups, the character does not capture all the information contained in the quantized theory.

The results for the unbounded sectors raise interesting questions about their interpretation in gravitational path integrals. Previous studies of the relation between the three-dimensional gravitational partition function and characters of coadjoint orbits have focused primarily on sectors with a Hamiltonian bounded from below, since saddles with an unbounded Hamiltonian are often regarded as unstable, and are excluded from the gravitational path integral due to the presence of negative modes. However, the existence of a finite path integral on orbits with an unbounded Hamiltonian suggests that the apparent instability of the associated gravitational saddles may likewise depend on the choice of integration contour. This viewpoint is consistent with recent studies of gravitational path integrals \cite{Marolf:2022ybi, Liu:2023jvm} based on Picard--Lefschetz theory \cite{Witten:2010cx, Witten:2010zr}.

Another important question is whether the results of path integral formalism on the orbits have higher-loop corrections. For the coadjoint orbits of Virasoro and BMS$_3$ with Hamiltonians bounded from below, previous studies \cite{Stanford:2017thb,Cotler:2018zff,Merbis:2019wgk,Cotler:2024cia, Simon:2024dwm} have established the one-loop exactness of the orbit path integral. Therefore, for these bounded orbits, the one-loop orbit path integrals computed here should be understood as exact quantum results rather than as merely leading-order semiclassical approximations. However, for the remaining orbits with Hamiltonians unbounded from below, the one-loop exactness has not yet been established. Investigating whether such orbits are likewise one-loop exact is therefore an interesting direction for future work.

Before closing, we comment on relations of the coadjoint modules constructed in the present work for the BMS$_3$ group and the HWRs. The latter arise naturally in the Carrollian CFT and have also played an important role in flat holography, see, e.g., in \cite{Hijano:2018nhq, Bagchi:2016bcd, Bagchi:2019unf}. More recent studies of Carrollian CFT provide explicit realizations of different vacuum \cite{Hao:2021urq, Hao:2022xhq, Chen:2024voz, Cotler:2024xhb, Chen:2025fcc}, while tensionless strings offer further examples of the existence of highest-weight vacua in Carrollian field theory \cite{Bagchi:2020fpr, Chen:2025gaz, Chen:2026cau}. In addition, it has been shown that the character for both induced representation and HWR can reproduce the gravitational partition function \cite{Chen:2025fcc}. However, agreement at the level of characters does not imply that the representations are identical. Actually, these two representations are structurally distinct, as we briefly compare below.

Generally, the reference state of HWR will be multiplet, but here we only compare the singlet with the induced representation, as the multiplet structure does not appear in the induced representation. The most familiar singlet highest-weight state is given by \cite{Bagchi:2019unf}
\begin{equation} \label{eq:BMSHWR}
    \begin{split}
        & L_0 \ket{\Delta, \xi} = \Delta \ket{\Delta, \xi}, \quad M_0 \ket{\Delta, \xi} = \xi \ket{\Delta, \xi}, \\
        & L_n \ket{\Delta, \xi} = M_n \ket{\Delta, \xi} = 0, \qquad (n \in \mathbb Z_+).
    \end{split}
\end{equation}
This can be obtained by taking an appropriate limit to HWRs of two copies of the Virasoro algebra $\{\mathcal{L}_n, c\} \oplus \{\bar{\mathcal{L}}_n, \bar c\}$. There are two ways to realize this, both are different from the limit used above to obtain the BMS$_3$ induced representations. If both Virasoro factors are standard HWR, the resulting BMS$_3$ HWR \eqref{eq:BMSHWR} is obtained by taking the Galilean limit rather than the UR limit \eqref{eq:ultraR} of two annihilation conditions of the Virasoro HWRs:
\begin{equation}
    (L_n, c_1) = (\mathcal{L}_n - \bar{\mathcal{L}}_n, c- \bar c), \qquad (M_n, c_2) = \epsilon (\mathcal{L}_n + \bar{\mathcal{L}}_n, c + \bar c).
\end{equation}
Alternatively, if one takes the UR limit \eqref{eq:ultraR}, the two Virasoro factors must have different highest-weight structures: the $\mathcal{L}_n$ sector obeys the standard highest-weight annihilation conditions \eqref{eq:Ln|0>Vir}, while the $\bar{\mathcal{L}}_n$ sector obeys the flipped highest-weight conditions:
\begin{equation}
    \begin{split}
        & \mathcal{L}_n \ket{b_0, \bar b_0} = \bar{\mathcal{L}}_{-n} \ket{b_0, \bar b_0} = 0, \qquad (n \in \mathbb Z_+), \\
        & \mathcal{L}_0 \ket{b_0, \bar b_0} = 2 \pi b_0 \ket{b_0, \bar b_0}, \quad \bar{\mathcal{L}}_0 \ket{b_0, \bar b_0} = 2 \pi \bar b_0 \ket{b_0, \bar b_0}.
    \end{split}
\end{equation}

Finally, we expect that the analysis presented here can be applied to other contexts. On the one hand, several recent works \cite{Sheikh-Jabbari:2026cnj, Sheikh-Jabbari:2026tpf, Sheikh-Jabbari:2026vqh} have studied extensions of the BMS$_3$ algebra by a weight-one operator. Constructing the corresponding coadjoint modules and computing their characters would provide a natural generalization of the present framework, and may also provide building blocks for understanding the quantized null string theory. Moreover, over the past decade, there has been a surge of interest in the ASG of four-dimensional spacetime at null infinity, namely the BMS$_4$ group \cite{Bondi:1962px,Sachs:1962wk,Sachs:1962zza}, which have been identified with soft gravitons \cite{He:2014laa,Strominger:2017zoo}. Subsequently, these soft modes in four dimensions have been studied in terms of unitary irreducible representation of the BMS$_4$ group \cite{Bekaert:2024uuy,Bekaert:2025kjb,Donnay:2026urd}. Moreover, recent work \cite{Henneaux:2018cst} suggested that the ASG of spatial infinity can also be BMS$_4$ by choosing proper boundary conditions. The local version of the BMS$_4$ group consists of the semi-direct product of supertranslations and two copies of Virasoro, the latter being referred to as superrotations \cite{Barnich:2009se,Barnich:2010eb,Barnich:2011mi,Barnich:2010ojg,Barnich:2017ubf}. Its coadjoint representation and geometric action were studied in \cite{Barnich:2021dta, Barnich:2022bni}, while further aspects of BMS$_4$ representations have been explored in \cite{Ruzziconi:2026isv}. It would therefore be interesting to determine whether the BMS$_4$ coadjoint-orbit can similarly describe boundary excitations of 4D flat gravity from the coadjoint modules constructed by the path integral on the orbits. We leave these questions for future work.

\acknowledgments

The authors would like to express their sincere gratitude to Prof. Bin Chen for his generous support and insightful suggestions on the manuscript. The authors are grateful to Stefan Prohazka for pointing out the subtleties associated with infrared divergences in the computation of the character. The authors thank Zezhou Hu, Zheng-Li Luo, Jun Nian, Blagoje Oblak, Yi-Nan Wang, Jie Xu, Zhenbin Yang, Zhi-jun Yin for other valuable discussions. P.M.~is supported in part by the National Natural Science Foundation of China (NSFC) under Grants No.~12475059 and No.~11935009, and by Tianjin University Self-Innovation Fund Extreme Basic Research Project Grant No.~2025XJ21-0007.

\appendix

\section{Central extension} \label{sec:central}

By definition, the representation $\mathcal{T}$ of a group $G$ is a map that preserves group multiplication, i.e., $\mathcal{T} [f] \mathcal{T} [g] = \mathcal{T} [fg]$. However, quantum states that differ only by a phase factor cannot be distinguished. This corresponds to the projective representation $\widetilde{\mathcal{T}}$, which does not strictly preserve group multiplication but instead introduces a phase factor $\widetilde{\mathcal{T}} [fg] = e^{i C(f,g)} \widetilde{\mathcal{T}} [f] \widetilde{\mathcal{T}} [g]$. The projective representation corresponds to the representation of the centrally extended group $\widehat{G} = G \ltimes \mathbb R$, endowed with the group multiplication
\begin{equation} \label{eq:multi}
    (g, \mu) (f, \lambda) = (gf, \lambda + \mu + C(g,f)),
\end{equation}
with a Lie group two-cocycle $C$. In particular, $C (f, f^{-1}) = C(1,g) = C(g,1)=0$. The inverse of a group element is $(f, \lambda)^{-1} = (f^{-1}, - \lambda)$. The representation in the new group $\widehat{G}$ corresponding to projective representation of the original group is denoted as $\widehat{\mathcal{T}}$, with $\widehat{\mathcal{T}}[(f, \lambda)] \widehat{\mathcal{T}}[(g, \mu)] = \widehat{\mathcal{T}}[(f, \lambda) (g, \mu)]$. The associative law yields that
\begin{equation} 
    C(f,gh) + C(g,h) = C(fg,h) + C (f,g).
\end{equation}

The Lie algebra of the centrally extended group $\widehat{G}$ is $\widehat{\mathscr G} = \mathscr G \oplus \mathbb R$, in which the Lie algebra element is derived as $(u,n) = \p_t (f_t, t n)|_{t = 0}$. Then, the finite and infinitesimal adjoint actions are then obtained as
\begin{align}
    \widehat{\text{Ad}}_g (u,n) & = \p_t \left[(g, \lambda) (f_t, t n) (g, \lambda)^{-1} \right] \big|_{t=0} = (\text{Ad}_g u, n - \<S(g), u\>_0), \label{eq:hatAddef} \\
    \widehat{\text{ad}}_{(u, \lambda)} (v, \mu) & = [(u,\lambda), (v, \mu)] = \p_t \left[ \widehat{\text{Ad}}_{(g_t, t \lambda)} (v, \mu) \right] \Big|_{t = 0} = ([u,v], \omega(u,v)), \label{eq:Liec}
\end{align}
where the Lie algebra two-cocycle $\omega(u,v) = - \<s (u), v\>_0$ is anti-symmetric, and the one-cocycles $S(g), s(u)$ are given by 
\begin{equation} \label{eq:CS}
    \<S(f), u\>_0 = -\p_t\left[ C (g_t, f^{-1}) + C (f, g_t f^{-1}) \right] \big|_{t=0}, \quad s (\p_t g_t |_{t=0}) = \p_t S (g_t) \big|_{t=0}\,,
\end{equation}
which is referred to as the Sourian construction. The Lie bracket can be rewritten as $[T_i, T_j] = f_{ij}^k T_k + \omega_{ij} \mathcal{Z}$. Here, $T_j$ serves as the basis of the adjoint representations such that the Lie algebra elements can be expanded as
\begin{equation}
    U = (u, n) = u^i T_i + n \mathcal{Z}, \qquad T_i = (t_i, 0), \quad \mathcal{Z} = (0, 1),
\end{equation}
where $\mathcal{Z}$ is the central charge of the algebra and $\omega_{ij} = \omega (t_i, t_j)$.

The dual of $\widehat{\mathscr G}$ is $\widehat{\mathscr G}^* = \mathscr G^* \oplus \mathbb R$. Then, the inner product between $(b, c) \in \widehat{\mathscr G}^*$ and $(u,n) \in \widehat{\mathscr G}$ is defined as
\begin{equation} \label{eq:innerc}
    \<(b, c), (u, n)\> = \<b, u\>_0 + cn.
\end{equation}
where $c$ is the central charge of the coadjoint vector and $\<\cdot, \cdot\>_0$ denotes the inner product between $\mathscr G$ and $\mathscr G^*$. The coadjoint actions can also be derived from the adjoint action \eqref{eq:hatAddef} with the help of inner product:
\begin{equation} \label{eq:Ad*gbc}
    \widehat{\text{Ad}}^*_g (b,c) = (\text{Ad}^*_g b - c S(g^{-1}), c), \quad \widehat{\text{ad}}^*_{(u, \lambda)} (b, c) =   (\text{ad}^*_u b + c s (u), 0),
\end{equation}
The coadjoint element can be expanded through the basis $(T^i)^*$ as 
\begin{equation}
    (b,c) = (T^i)^* b_i + c \mathcal{Z}^*, \qquad (T^i)^* = ((t^i)^*,0), \quad \mathcal{Z}^* = (0,1).
\end{equation}
where $\<(T^i)^*, T_j\> = \delta^i_j, ~ \<\mathcal{Z}^*, \mathcal{Z}\> = 1$. The symplectic form is then derived as 
\begin{equation}
    \begin{split}
        \Omega_b \left( \widehat{\text{ad}}^*_V (b, c), \widehat{\text{ad}}^*_{V'} (b, c) \right) = \< (b, c), [V, V'] \> = v^i v'^j \Omega_{b, ij} = v^i v'^j (b_k f^k_{ij} + c \omega_{ij}).
    \end{split}
\end{equation}
Therefore, the Poisson bracket $\{b_i, b_j \} = \Omega_{b, ij}$ also reproduces the Lie algebra with the central charge $\mathcal{Z}$ replaced by $c$.

The conserved charges are modified by 
\begin{equation} \label{eq:chargec}
    \Phi_v (b) = \<(b,c), (v, n)\> = b_i v^i +cn
\end{equation}
as in \eqref{eq:charge}, and one can also verify that 
\begin{equation} \label{eq:PoissonPhi}
    \{\Phi_v, \Phi_{v'}\} = \Phi_{[v, v']} + c \omega(v, v') = \Omega_b \left( \widehat{\text{ad}}^*_V (b, c), \widehat{\text{ad}}^*_{V'} (b, c) \right).
\end{equation}

Similarly, each point $(b,c)$ on the orbits $W_{(b_0, c)}$ can be expressed by the reference point $(b_0, c)$ through a finite transformation $(b, c) = \widehat{\text{Ad}}^*_g (b_0, c)$, and satisfies the MC equation $\delta (b, c) = \widehat{\text{ad}}^*_U (b, c)$ with the MC form defined in the same way as in \eqref{eq:delub}
\begin{equation} \label{eq:umu}
    U = (u, m_u) = \delta (g, \mu) (g, \mu)^{-1} = \left( \delta g g^{-1}, \delta \mu + [\delta_g C (g, f)] |_{f=g^{-1}} \right),
\end{equation}
where we applied the group multiplication in \eqref{eq:multi}.

The geometric action after the central extension can be derived by the inner product between the coadjoint vector $(b, c)$ and its corresponding MC form $U$
\begin{equation} \label{eq:geoc}
    A = -\int \<(b, c), U \> = \int (\alpha_b + \alpha_c), \quad \alpha_c = -c m_u + c \< S (g^{-1}), \delta g g^{-1}\>_0\,,
\end{equation}
where $\alpha_b$ is given in \eqref{eq:Omega}, and $A$ can also be expressed as $\int \td t p_s \dot q_s$ with the Darboux coordinates $(p_s, q_s)$ modified by the central charge.


\providecommand{\href}[2]{#2}\begingroup\raggedright\endgroup

\end{document}